\documentclass[sensors,review,accept,moreauthors,pdftex,10pt,a4paper]{mdpi} 
\firstpage{1} %%MDPI internal note: new layout%%
\articlenumber{0000}
\doinum{10.3390/------}
\pubvolume{16}
\pubyear{2016}
\copyrightyear{2016}
\externaleditor{Academic Editor: Kemal Akkaya}
\history{Received: 29 July 2016; Accepted: 29 September 2016; Published: date}
\usepackage{longtable}
\usepackage{url}
\usepackage[utf8]{inputenc}

\usepackage{multirow}
\usepackage{acronym}
\newacro{COTS}[COTS]{Commercial Off-The-Shelf}
\newacro{WSN}[WSN]{Wireless Sensor Networks}
\newacro{BLE}[BLE]{Bluetooth Low Energy}
\newacro{COTS}[COTS]{Commercial Off-The-Shelf}
\newacro{HTTP}[HTTP]{Hyper Text Transfer Protocol}
\newacro{SPI}[SPI]{Serial Peripheral Interface}
\newacro{UPB}[UPB]{Universal Powerline Bus}
\newacro{BT}[BT]{Bluetooth}
\newacro{RSSI}[RSSI]{Received Signal Strength Indicator}
\newacro{DoD}[DoD]{Department of Defense}
\newacro{AAA}[AAA]{Authentication, Authorization, and Accounting}
\newacro{ACV}[ACV]{Armored Combat Vehicles}
\newacro{AJ}[AJ]{Anti-Jamming} 
\newacro{AMC}[AMC]{Adaptive Modulation and Coding}
\newacro{AMC2}[AMC]{Adaptive Modulation Control}
\newacro{API}[API]{Application Programming Interface}
\newacro{ARQ}[ARQ]{Automatic Repeat Request}
\newacro{BFT}[BFT]{Blue Force Tracking}
\newacro{BS}[BS]{Base Station}
\newacro{C2}[C2]{Command and Control}
\newacro{C4I}[C4I]{Command, Control, Communications, Computers, and Intelligence}
\newacro{C4ISR}[C4ISR]{Command, Control, Communications, Computers, Intelligence, Surveillance and Reconnaissance}
\newacro{CC}[CC]{Command Center}
\newacro{COTS}[COTS]{Commercial Off-The-Shelf}
\newacro{CP}[CP]{Command Post}
\newacro{CR}[CR]{Cognitive Radio }
\newacro{DOA}[DOA]{Direction Of Arrival}
\newacro{DSS}[DSS]{Data Distribution Subsystems}
\newacro{ECRTP}[ECRTP]{Enhanced Compressed Real Time Protocol}
\newacro{EMCON}[EMCON]{Emissions Control}
\newacro{EPM}[EPM]{Electronic Protection Measures}
\newacro{IMS}[IMS]{IP Multimedia Subsystem}
\newacro{IST}[IST]{Information Systems Technology Panel}
\newacro{JIE}[JIE]{Joint Information Environment}
\newacro{LMR}[LMR]{Land Mobile radio}
\newacro{LOS}[LOS]{Line Of Sight}
\newacro{LPD}[LPD]{Low Probability of Detection}
\newacro{LPI}[LPI]{Low Probability of Interception}
\newacro{LTE}[LTE]{Long-Term Evolution}
\newacro{MBWCS}[MBWCS]{Military Broadband Wireless Communication System}
\newacro{MPE}[MPE]{Mission Partner Environment}
\newacro{NCW}[NCW]{Network Centric Warfare}
\newacro{NII}[NII]{Network Information Infrastructure}
\newacro{NLOS}[NLOS]{Non Line-Of-Sight}
\newacro{NNEC}[NNEC]{NATO Network Enabled Capability}
\newacro{OTA}[OTA]{Over-The-Air}
\newacro{OTAR}[OTAR]{Over-The-Air Rekeying}
\newacro{PHS}[PHS]{Packet Header Suppression}
\newacro{PMP}[PMP]{Point-To-Multipoint}
\newacro{PtP}[PtP]{Point-to-Point}
\newacro{PS}[PS]{Public Safety}
\newacro{PTT}[PTT]{Push-To-Talk}
\newacro{QoS}[QoS]{Quality of Service}
\newacro{ROHC}[ROHC]{Robust Header Compression}
\newacro{RTG}[RTG]{Research Task Group}
\newacro{SDR}[SDR]{Software Defined Radio}
\newacro{SIP}[SIP]{Session Initiation Protocol}
\newacro{Wi-Fi}[Wi-Fi]{Wireless Fidelity}
\newacro{WiMAX}[WiMAX]{Worldwide Interoperability for Microwave Access}
\newacro{WLAN}[WLAN]{Wireless Local Area Network}
\newacro{RFID}[RFID]{Radio-frequency identification}
\newacro{PaaS}[PaaS]{Platform as a Service} 
\newacro{UAV}[UAV]{Unmanned Aerial Vehicle}
\newacro{COP}[COP]{Common Operational Picture}
\newacro{QoI}[QoI]{Quality of Information}
\newacro{VoI}[VoI]{Value of Information}
\newacro{IETF}[IETF]{Internet Engineering Task Force}
\newacro{IEEE}[IEEE]{Institute of Electrical and Electronics Engineers}
\newacro{ETSI}[ETSI]{European Telecommunications Standards Institute}
\newacro{SOA}[SOA]{Service-Oriented Architectures}
\newacro{NFC}[NFC]{Near Field Communication}
\newacro{ISR}[ISR]{Intelligence Surveillance and Reconnaissance}

\usepackage{soul}
\usepackage{upgreek}
\usepackage{enumitem}
\usepackage{subfigure}
\usepackage{booktabs}
\usepackage{multirow}
\usepackage{longtable}
\usepackage{microtype}
\newcommand\myurl[1]{\changeurlcolor{black}\url{#1}\changeurlcolor{blue}}

\usepackage{subfigure}
\makeatletter
\renewcommand{\@thesubfigure}{\normalsize(\textbf{\alph{subfigure}})}
\makeatother

\usepackage{supertabular}
\usepackage{afterpage}
\Title{A Review on Internet of Things for Defense and Public Safety} 
\Author{Paula Fraga-Lamas*, Tiago M. Fern\'andez-Caram\'es, Manuel Su\'arez-Albela, Luis Castedo and \mbox{Miguel Gonz\'alez-L\'opez}}
\address[1]{ Department Electronics and Systems, Faculty of Computer Science, Universidade da Coru\~na, A~Coru\~na, 15071, Spain;
tiago.fernandez@udc.es (T.M.F.-C.); m.albela@udc.es (M.S.-A.); luis.castedo@udc.es~(L.C.); miguel.gonzalez.lopez@udc.es (M.G.-L.)}

\corres{\hspace{-1em}Correspondence: paula.fraga@udc.es; Tel.: +34-9-8116-7000 (ext. 6051)}

\abstract{The Internet of Things (IoT) is undeniably transforming the way that organizations communicate and organize everyday businesses and industrial procedures. Its adoption has proven well suited for sectors that manage a large number of assets and coordinate complex and distributed processes. This survey analyzes the great potential for applying IoT technologies (i.e., data-driven applications or embedded automation and intelligent adaptive systems) to revolutionize modern warfare and provide benefits similar to those in industry.
It identifies scenarios where Defense and Public Safety (PS) could leverage better commercial IoT capabilities to deliver greater survivability to the warfighter or first responders, while
reducing costs and increasing operation efficiency and effectiveness. This article reviews the main tactical requirements and the architecture, examining gaps and shortcomings in existing IoT systems across the military field and mission-critical scenarios. The review characterizes the open challenges for a broad deployment and presents a research roadmap for enabling an affordable IoT for defense and PS.
}


\begin{document}

\section{Introduction}
The Internet of Things (IoT) is a distributed system for creating value out of data. It enables heterogeneous physical objects to share information and coordinate decisions.
The impact of IoT in the commercial sector results in significant improvements in efficiency, productivity, profitability, decision-making and effectiveness. IoT is transforming how products and services are developed and distributed, and how infrastructures are managed and maintained. It is also redefining the interaction between people and machines. From energy monitoring on a factory \cite{energyFactory} to tracking supply chains~\cite{SupplyChain}, IoT optimizes the performance of the equipment and enhances the safety of workers.
Until~today, it has allowed for more effective monitoring and coordination of manufacturing, supply chains, transportation systems, healthcare, infrastructure, security, operations, and industrial automation, among other sectors and processes.

\begin{figure}[H]
\centering
\includegraphics[width=1\columnwidth]{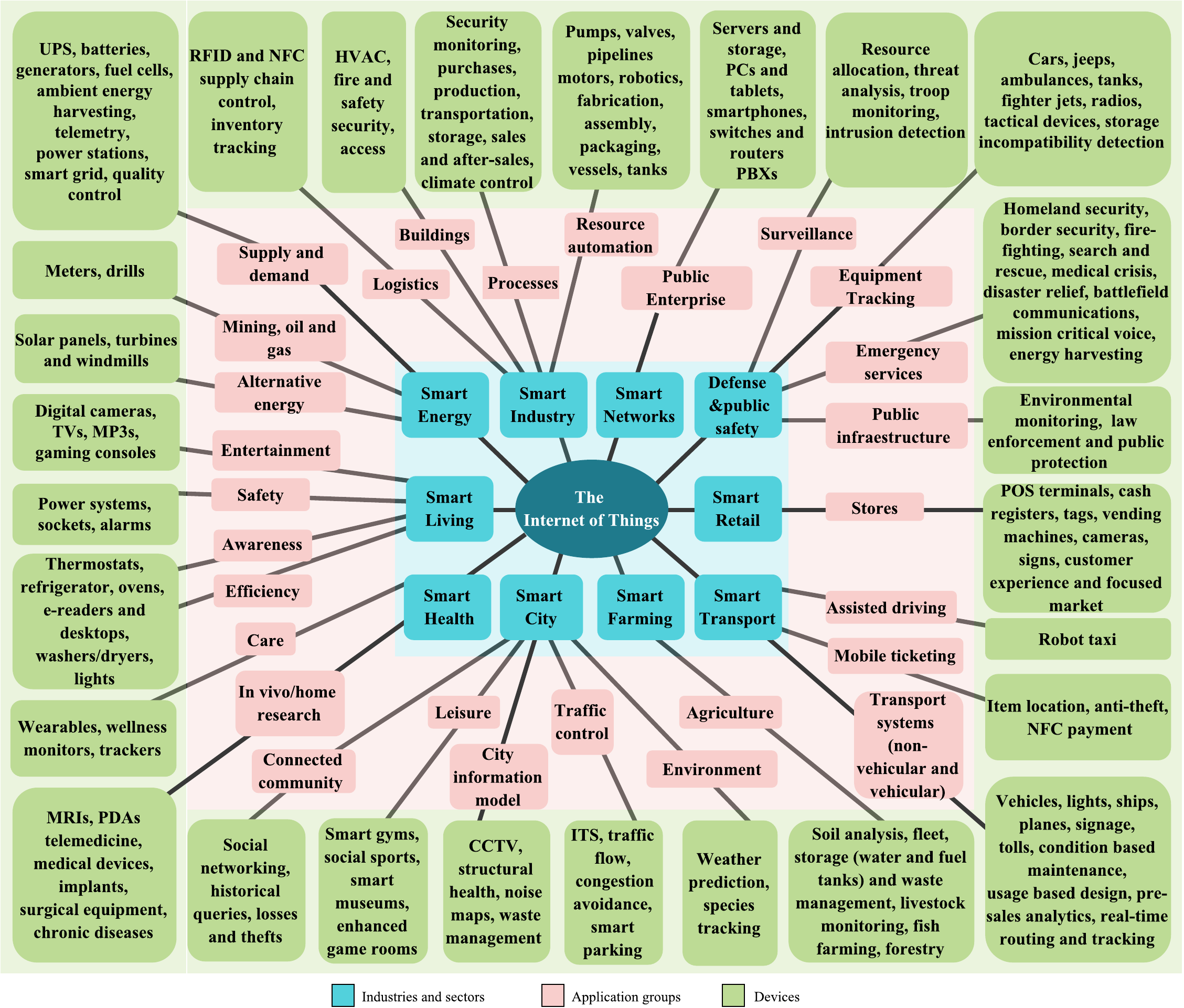}
\caption{Proliferation of devices and applications in the Internet of Things (IoT).}
\label{fig:ProliferationDevices}
\end{figure}

IoT is estimated to reach 50 billion connected devices by 2020 and the potential economic impact will be from \$3.9 trillion to \$11.1 trillion per year by 2025 \cite{IoTMcKinsey}. Overall, IoT would allow for the automation of everything around us.
The proliferation of devices and its applications is illustrated in Figure \ref{fig:ProliferationDevices}.

Regarding Machine-to-Machine (M2M) communications, traffic volume is expected to increase at an annual growth rate of 25 percent up to 2021. In total, in such a year there will be around 28 billion connected devices with more than 13.2 billion using M2M communications \cite{Ericsson}. 

Currently, the industrial and business sector is leading the adoption of IoT.  Businesses will spend \$3 billion in the IoT ecosystem and deploy 11.2 billion devices by 2020, while customers will  invest up to \$900 million \cite{BusinessInsider}. On the other hand, the public sector is estimated to increase significantly its adoption and spend up to \$2.1 billion and install 7.7 billion devices, being the second-largest adopter of IoT ecosystems, particularly in areas like smart cities \cite{Zanella, Vermesan}, energy management \cite{energy} and transportation~\cite{IoV}.

IoT represents the convergence of several interdisciplinary domains  \cite{leveragingIoT, Al-Fuqaha, Miorandi, Atzori, nanothings}:
networking, embedded hardware, radio spectrum, mobile computing, communication technologies, software architectures, sensing technologies, energy efficiency, information management, and data analytics. The rapid growth of IoT is driven by four key advances in digital technologies. 
The first one is the declining cost and miniaturization of ever more powerful microelectronics such as transducers (sensors and actuators), 
processing units (e.g., microcontrollers, microprocessors, SOCs (System-on-a-chip), FPGAs (Field-Programmable Gate Array), and receivers. The second factor is the fast pace and expansion of wireless connectivity.
The third is the expansion of data storage and the processing capacity of computational systems.
Finally, the fourth one is the advent of innovative software applications and analytics, including advancements in machine-learning techniques for big data processing.
These four drivers are present in the layers of the IoT technology stack. 
For instance, IoT~may include transducers that collect data on physical and environmental conditions. These devices transmit data over a wired or wireless communication network to servers and computers that store and process data using software applications and analytics. The knowledge gleaned from the analysis can be used for fault detection, control, prediction, monitoring, and optimization of processes and~systems. 

% Unique capabilities prior invented by defense departments can now be purchased as Commercial Off-The-Shelf (COTS) components. In the 50s, the American \ac{DoD} played a critical role in pioneering the sensor, lightweight computer networking and communications technology. That served as the foundation for IoT technologies as we know today. For example, wireless sensor network technologies were first developed by the U.S. military to detect and track Soviet submarines \cite{SOSUS}. 

As it will be seen in Section \ref{ScenariosIoT}, IoT technologies have the potential to increase tactical efficiency, effectiveness, safety and deliver immense cost savings in the long-term. These technologies can help the military and first responders to adapt to a modern world in which adversaries are located in more sophisticated and complex suburban scenarios (smart cities) while budgets are shrinking.

Defense and Public Safety (PS) organizations play a critical societal role ensuring national security and responding to emergency events and catastrophic disasters.  
Instead of PS, some authors use the term Public Protection Disaster Relief (PPDR) \cite{PPDR} radio communications, defined in ITU-R
%Please define.
 Resolution 646 (WRC-12) as a combination of two key areas in emergency response:
\begin{itemize}[leftmargin=*,labelsep=5mm]
\item Public protection (PP) radio communication: communications used by agencies and
organizations responsible for dealing with the maintenance of law and order, protection of life
and property, and emergency situations.
\item Disaster relief (DR) radio communication: communications used by agencies and
organizations dealing with a serious disruption in the functioning of society, posing a
significant, widespread threat to human life, health, property or the environment, whether
caused by accident, nature or human activity, and whether they happen suddenly or as a result of complex, long-term processes.
\end{itemize}

Nowadays, the  challenge of crisis management is in reducing the impact and injury to individuals and assets. This task demands a set of capabilities previously indicated by European TETRA~\cite{TETRA}, TCCA~\cite{TCCA}, and ETSI \cite{ETSI} standardization bodies and American APCO Project-25 \cite{APCO}, 
%Please provide definitions here.
which~includes resource and supply chain management, access to a wider range of information and secure communications. Military and first responders should be able to exchange information in a timely manner to coordinate the relief efforts and to develop situational awareness. FY 2016 SAFECOM Guidance \cite{SAFECOM} provides an overview of emergency communications systems and technical standards. Communication capabilities need to be provided in very challenging environments where critical infrastructures are often degraded or destroyed. Furthermore, catastrophes, natural disasters or other emergencies are usually unplanned events, causing panic conditions in the civilian population and affecting existing resources. In large-scale natural disasters, many different PS organizations (military organizations, volunteer groups, non-government organizations and other local and national organizations) may be involved. At the same time, commercial communication infrastructure and resources must also be functional in order to alert and communicate with the civilian population. In~addition, specific security requirements including communication and information protection can also exacerbate the lack of interoperability. Sharing various types of data is needed in order to establish and maintain a Common Operational Picture (COP) between agencies and between field and central command staff. %; and the provision of wireless broadband communication requires the availability of radio frequency spectrum bands.  

Typically, first responders include police officers, firefighters, border guards, coastal guards, emergency medical personnel, non-governmental organizations (NGOs) and other organizations among the first on the scene of a critical situation. These organizations can provide one or more of the functions described above. The relationships between them may depend on the national legislation or the context. Table \ref{table:PS} provides an overview of the main functions and the potential relationships among the agents of the different PS organizations.

The rest of this article we will be focus on the military side, since we consider that it covers most of the significant scenarios and functions, and it represents the most challenging cases (in terms of operational and technical requirements) in PS organizations.
 
\newpage

 \begin{center}
 \begin{longtable}{p{2.47cm}p{12.45cm}c} 
 
\caption{{Public Safety (PS) agents.}}\label{table:PS}\\ 

 \toprule \multicolumn{1}{c}{\textbf{PS Organizations}} & \multicolumn{1}{c}{\textbf{Description}}   \\ \midrule

%\endfirsthead
%\hline
%\multicolumn{2}{c}%
%{{\bfseries \tablename\ \thetable{} -- continued from previous page}} \\
%\hline \multicolumn{1}{|c|}{\textbf{PS Organizations}} & \multicolumn{1}{c|}{\textbf{Description}}  \\ \hline 
%\endhead
%
%\hline \multicolumn{3}{|r|}{{Continued on next page}} \\ \hline
%\endfoot

%\hline \hline
%\endlastfoot

\multirow{4}{*}{Police officers}
& $\bullet$ Law enforcement and protection of the citizens: keep the peace and secure volatile areas, prevent and investigate crime, detain individuals suspected/convicted of offenses against criminal law.\\
& $\bullet$ Urban/rural environments, major events and border areas.\\
\midrule

\multirow{6}{*}{Fire services}
& $\bullet$ Law enforcement, protection of the environment, search and rescue. With~variations from country to country, the primary areas of responsibility include: structure fire-fighting and fire safety, wild land fire-fighting, life-saving through search and rescue, rendering humanitarian services, management of hazardous materials, environment protection, salvage and damage control, safety~management
within an inner cordon and mass decontamination.\\ 
& $\bullet$ Urban/rural environments, ports and airports.\\ 
\midrule

\multirow{4}{*}{Border guards}
& $\bullet$ National security agencies which perform land border control against criminal activities and control of illegal immigration. They can also be involved in cross-national disaster management.\\ 
& $\bullet$ Rural environments and border security areas (green border).\\
\midrule

\multirow{6}{*}{Coastal guards}
& $\bullet$ Law enforcement, protection of the environment, search and rescue (at~sea~and other waterways), protection of coastal waters, criminal interdiction, illegal~immigration, disaster and humanitarian assistance in areas of operation. They~can also be involved in cross-national disaster management (e.g.,~earthquake,~flooding).\\ 
& $\bullet$ Border areas (Blue border) and ports.\\  
\midrule

\multirow{3}{*}{Medical responders}
& $\bullet$ Emergency medical services provide critical care of sick and injured
citizens, and the ability to transfer the people in a safe and controlled environment. \\ 
& $\bullet$ All scenarios.\\
\midrule

\multirow{3}{*}{Road agents}
& $\bullet$ Police agency responsible for the law enforcement and protection of road transportation
ways.\\ 
& $\bullet$ Urban/rural environments.\\
\midrule

\multirow{3}{*}{Railway agents}
& $\bullet$ Specialized police agency responsible for the protection and law enforcement of road transportation.\\ 
& $\bullet$ Urban/rural environments.\\
\midrule

%\pagebreak
\multirow{3}{*}{Custom guards}
& $\bullet$ Law enforcement, crime prevention, monitoring people and goods entering
a~country.\\ 
& $\bullet$ Border areas.\\
\midrule

\multirow{2}{*}{Airport security}
& $\bullet$ Law enforcement, protect airports, passengers and aircrafts from crime.\\ 
& $\bullet$ Urban/rural environments.\\
\midrule

\multirow{5}{*}{Military}
& $\bullet$ It is the organization responsible for the national defense policy and the protection of national security. It may also supports PS organizations in case of a large national disaster, terrorist acts, major events, search and rescue or emergency medical services.\\ 
& $\bullet$ All scenarios.\\
\bottomrule

\end{longtable}
\end{center}

Over the last years, some research papers focused on evolving PS organizations have been published \cite{earthquake, evolvingPS,PSstudy,JTRS,PSsurvey}. Some of these articles have particular interest in the challenges to evolve the LTE network architecture toward 5G in order to support emerging PS networks \cite{PSnetworks, PSnetworks2, PSnetworks3}.
With respect to IoT, there are several published papers that cover different aspects of the IoT technology applied to defense and PS. For example, Chudzikiewicz et al.  \cite{Zielinski_NATO} propose a fault detection method that is based on a network partitioned into clusters for the military domain. Yushi et al. \cite{MIOT} introduce a layer architecture and review some application modes. They also include the example of a weapon control application. Butun et al. \cite{PScloud} propose a lightweight, cloud-centric, multi-level authentication as a service approach that addresses scalability and time constraints for IoT devices surrounding PS responders.
References \cite{leveragingIoT, Wind, Deloitte, Survey_ICMCIS2016} contain short surveys for leveraging the IoT for a more efficient military. The authors of \cite{Wrona, SecurityIoT, Algassem} focus on security challenges, while TCG drafts a guideline for securing IoT networks \cite{TCG}.

Unlike recent literature, the contribution of this article focuses on providing a holistic approach to IoT applied to defense and public safety (PS) with a deeper study of the most relevant operational requirements for mission-critical operations and defense, an overview of the key challenges, and the relationship between IoT and other emerging technologies. Besides, the aim of this article is to help the defense industry to exploit the opportunities created by using commercial IoT applications in mission-critical environments.

The remainder of this paper is organized as follows.
Section \ref{SecStateIoT} analyzes the state of the art of current \ac{COTS} technologies and its application for defense and PS.
Section~\ref{ScenariosIoT} presents some promising scenarios for mission-critical IoT.
Section \ref{Requirements} introduces the main operative requirements and capabilities, and analyzes their applicability to defense and PS.
Section \ref{BasicsIoT} reviews the basics of the IoT architecture for tactical and emergency environments.
Section \ref{Challenges}  describes the main shortcomings, and outlines the primary technical and cultural challenges that stand in the way of leveraging IoT technologies at a broader scale. It also identifies further research areas to enable COTS IoT for tactical and emergency environments.
Finally, Section \ref{SecCon} is devoted to conclusions.

\section{Compelling COTS IoT Applications}\label{SecStateIoT}

The introduction of the Network-Centric Warfare (NCW) paradigm \cite{NCW2015,NCW2} transformed traditional military approaches by reversing the policy of expanded communication gateways, and by connecting battlefield assets back to headquarters. Through sharing data between both legacy assets and new deployments, this military vision creates advantages through force projection and secure timely exchanges of information between users. The NCW paradigm integrates three domains: 
the physical domain, generating data where events happen and operations are conducted;
the information domain, in which data are transmitted and stored;
and the cognitive domain, in which data are processed and analyzed supporting decision-making.
The three domains of NCW translate directly into the foundations of today's commercial IoT.

Consequently, this driving paradigm has led to the adoption of IoT-related technologies in key areas. Today, military and homeland security are struggling to equip its workforce with the basic functions provided by COTS technologies, i.e., commercial smart phones or \ac{RFID}. Indeed, defense continues to drive innovation regarding advanced sensors, surveillance and reconnaissance drones, satellite communication systems, and control systems, 
and has invested significantly in mobile technologies, including tactical mobility for warfighters.

\begin{figure}[H]
\centering
\includegraphics[width=1\columnwidth]{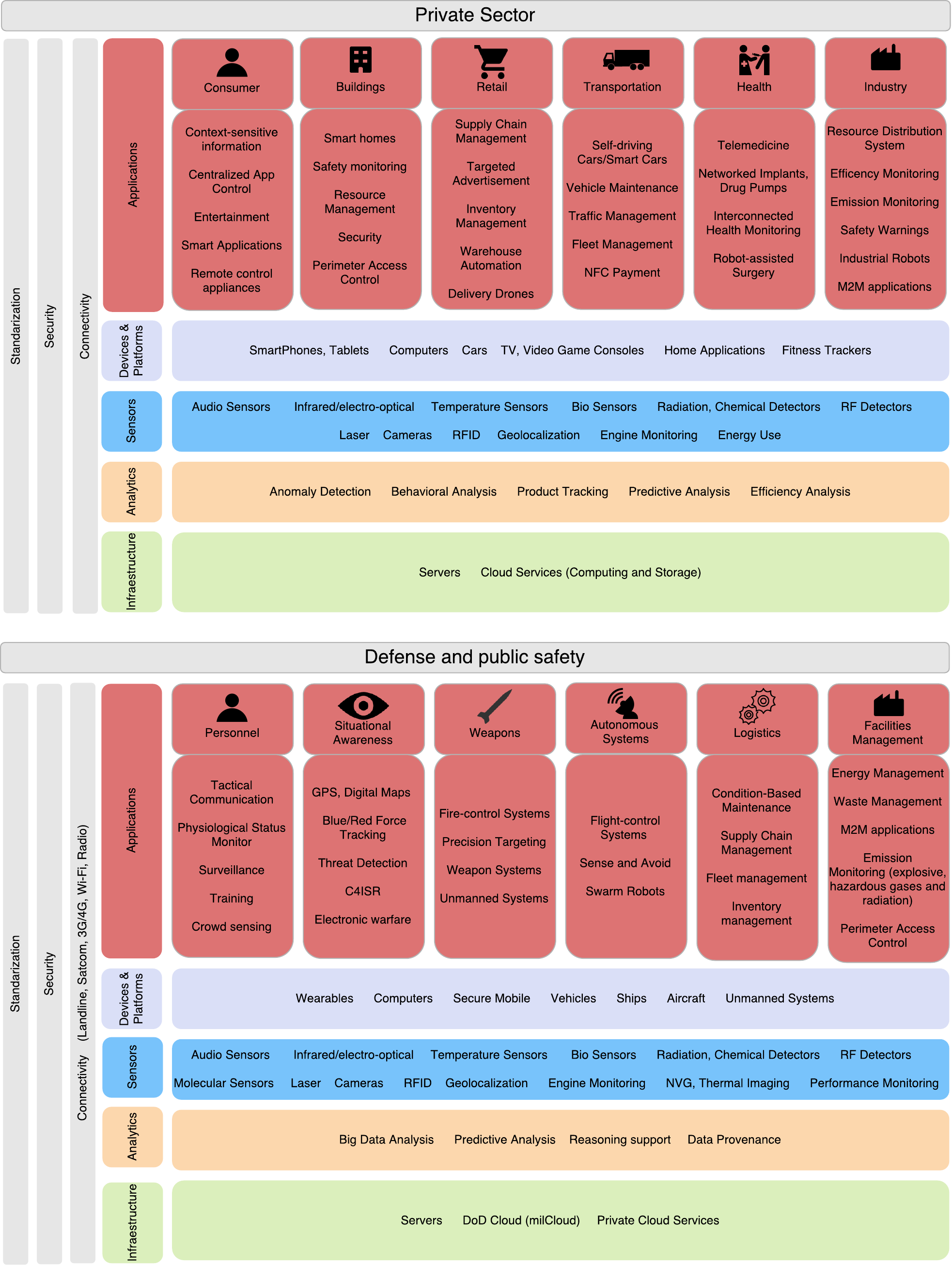}
\caption{Private sector vs defense and public safety technology stack.}
\label{fig:PrivateSectorVsMilitaryTechnologyStack}
\end{figure}

Nevertheless, the vast majority of current IoT applications are driven by the private sector, with the military severely lagging behind in everyday operations. Defense has today an opportunity to seize benefits from IoT by partnering with the private sector and adopting IoT-enabled business practices that consider the technological particularities of tactical systems. Figure \ref{fig:PrivateSectorVsMilitaryTechnologyStack}
shows a comparison between the private sector and the defense and public safety technology stack. 

\afterpage{\clearpage}

This section provides an overview of COTS IoT applications that might be relevant and enlightening to defense and public safety environments. We have to remark the distinction between ad-hoc IoT military deployments and already established civil deployments that have to be protected. 
Examples of IoT deployment are numerous \cite{Survey_ICMCIS2016}, but 
the following cases are illustrative of 
areas where the civil deployment has achieved significant benefits that can be leveraged by the military. They~include transportation, energy efficiency, inventory managements, or mining.  
\begin{itemize}[leftmargin=*,labelsep=5mm]

\item Transportation: Telogis \cite{Telogis}, which builds engine-monitoring systems for General Motors (GM) vehicles, estimates that its smart engine reduces fuel consumption by 25\%, around 30\% in idle time, increases fleet use by 25\%, and workforce productivity by 15\%.
 
\item Energy efficiency: IoT-based energy management systems can reduce energy use in office by 20\%~\cite{IoTMcKinsey}. Smart thermostats and HVAC (Heating, Ventilation and Air Conditioning) save consumers as much as 10\%--15\% on heating and cooling. Additionally, smart appliances for home automation systems \cite{suarez2016} are now being researched \cite{fernandez2015intelligent}. IoT-based use cases for smart cities have been also identified, like traffic monitoring \cite{TrafficMonitoring}, surveillance \cite{surveillance} and pollution monitoring \cite{pollution}. Within this scenario, the content of the samples acquired may expose critical information. For example, in the case of noise pollution monitoring applications, the noise may also contain private conversations.

\item Inventory management: USTRANSCOM's Global Transportation Network (GTN) \cite{USTRANSCOM} and DLA (Defense Logistics Agency) developed a common information platform that enables the military to improve end-to-end supply visibility, service and logistics processes.  The platform includes a single repository and universal access to logistics data so that any user or developer can easily access or manage supply chain information. Also, such a platform facilitates the development of new applications that run on the same backbone. Another contribution of USTRANSCOM is fleet management, which uses RFID trackers to monitor palletized shipments among major transit~hubs.

\item Mining: embedded sensors on equipment and vehicles offer a more precise picture of ground operations and enable real-time monitoring of equipment. This new technology, together~with autonomous mining systems, is transforming daily operations. An IoT deployment reduces expenditure on infrastructure and machines by reducing outages and maintenance, energy~consumption and environmental impact, while significantly improving productivity and  mine safety by reducing injuries and fatalities. A renowned example is the autonomous drilling system of Rio Tinto (Perth, Australia), which includes tunneling machines, trains, autonomous haulage systems, and driverless trucks. This mission-control site manages operations of 15 mines, 31 pits, 4 port terminals  and a 1600\,km rail network \cite{RioTinto}.
\end{itemize}

The above-mentioned examples illustrate some of the ways in which the private sector is leveraging IoT and
enabling new business models. Today, European research projects like RERUM~\cite{RERUM}, RELYonIt \cite{RELYonIt}, FIESTA-IoT \cite{FIESTA-IoT}, BIGIoT \cite{BIGIoT}, or bloTope \cite{bloTope} intend to tackle the most relevant shortcomings of commercial IoT. Among the different projects, EU-funded METIS-II \cite{METISII} must be highlighted: the researchers created a 5G radio-access network design and provided the technical enablers needed for an efficient integration. Other concerns, like environmental issues or green deployment schemes for IoT \cite{GreenSw}, still remain open.

\section{Target Scenarios for Mission-Critical IoT} \label{ScenariosIoT}
\begin{figure}[H]
\centering
\includegraphics[scale=0.7]{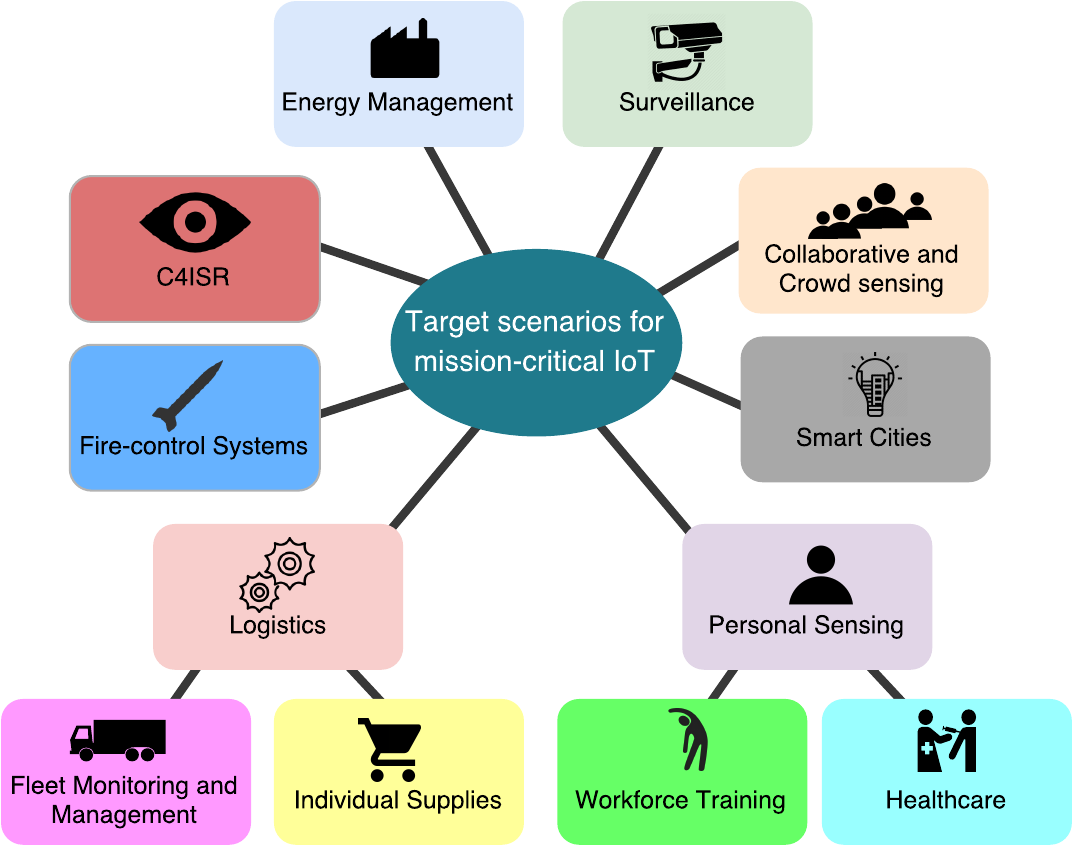}
\caption{Promising target scenarios for defense and public safety.}
\label{fig:TargetScenarios}
\end{figure}

In this section we provide an overview of the most promising IoT scenarios, 
which are depicted in Figure \ref{fig:TargetScenarios}. Until now, the deployment of IoT-related technologies for defense and public safety has been essentially focused on applications for \ac{C4ISR}, and fire-control systems. 
This is driven by a predominant view that sensors serve foremost as tools to gather and share data, and create a more effective Command and Control (C2) of assets.
IoT technologies have also been adopted in some applications for logistics and training, but their deployment is limited and poorly integrated with other systems.
%\afterpage{\clearpage}
Besides, IoT functionalities are useful for establishing advanced situational awareness in the area of operations. Commanders make decisions based on real-time analysis generated by integrating data from unmanned sensors and reports from the field. These commanders benefit from a wide range of information supplied by sensors and cameras mounted on the ground, and manned or unmanned vehicles or soldiers. These devices examine the mission landscape and feed data to a forward base. Some of the data may be relayed to a Command Center where it is integrated with data from other~sources.

\subsection{C4ISR}

\ac{C4ISR} systems use millions of sensors deployed on a range of platforms to provide advanced situational awareness.
Radar, video, infrared or passive RF detection data are gathered by surveillance satellites, airborne platforms, UAVs (Unmanned Aerial Vehicles), ground stations, and soldiers in the field.  These data are delivered to an integration platform that analyzes them and delivers information up and down the chain of command. These platforms provide a \ac{COP} allowing for enhanced coordination and control across the field.

High-level military echelons are provided with comprehensive situational awareness through central operations centers, which receive data feeds from platforms. Lower levels (i.e., platoon, soldiers) also have access to the data in their area. In the case of combat pilots, they receive prioritized data feeds integrated with data from their own sensor systems.

\subsection{Fire-Control Systems}
In fire-control systems, end-to-end deployment of sensor networks and digital analytics 
enable fully automated responses to real-time threats, and deliver firepower with pinpoint precision. For~example, the U.S. Navy's Aegis Combat system provides C2 as well as an unprecedented ballistic missile defense \cite{Aegis}. Munitions can also be networked, allowing smart weapons to track mobile targets or be redirected in flight. 
Prime examples are the Tomahawk Land Attack Missile (TLAM) and its variants, navy's precision strike standoff weapons for attack of long range, medium range and tactical targets \cite{Tomahawk}. Furthermore, the military has invested in the use of long endurance UAVs to engage high-value targets and introduce multi-UAVs applications \cite{UAV2}.

\subsection{Logistics}

Logistics is an area where multiple low-level sensors are already being used in defense. Currently, their deployment remains constrained to benign environments with infrastructure and human involvement. The military has already deployed some IoT technologies in non-combat scenarios in order to improve back-end processes. 
For example, RFID tags have been used to track shipments and manage inventories between central logistics hubs.

In the following subsections, we describe examples that belong to two main categories: fleet management and individual supplies. 

\subsubsection{Fleet Monitoring and Management}

Fleet monitoring can be represented by aircraft and ground vehicle fleets with on-board sensors that monitor performance and part status.  For example, they track vehicle status and subsystems, and indicate when resupplying low-stock items (i.e., fuel or oil) is needed. Sensors would issue alerts, potentially reducing the risk of fatal failures. The aim is to facilitate condition-based maintenance and on-demand ordering of parts, reduce maintenance staff, and decrease unanticipated failures or unnecessary part replacements.
Although IoT deployment carries up-front costs, it can enable significant long-term savings by transforming business processes across logistics.
Defense has an opportunity to take advantage in the auto and industrial sectors, and exploit performance data on existing data links, like Blue Force Tracker transponders (already in place on many military vehicles) to limit new security risks. By extension, IoT-connected vehicles could also share information, for~example, about available spare parts.

Real-time fleet management includes geolocation, status monitoring, speed and engine status, total engine hours, fuel efficiency, and weight and cargo sensors. Besides, when tracking shipments, the position and status of the containers can be monitored to identify potential problems.

Regarding aircraft, modern jet engines are equipped with sensors that produce several terabytes of data per flight. This information combined with in-flight data can improve engine performance to reduce fuel costs, detect minor faults or shorten travel duration \cite{MorganStanley}. Furthermore, it enables preventive maintenance resulting in a long lifecycle (slowing or preventing breakage) and less downtime spent in repairs.
The flight data can be tracked in real-time by operators and analysts on the ground.

\subsubsection{Individual Supplies}
The deployment of RFID tags, sensors and standardized barcodes allow for tracking individual supplies. 
IoT provides real-time supply chain visibility (whether it is being shipped, transferred, deployed, consumed...), and allows the military to order supplies on demand and simplify logistics management for operational units.
This smarter procurement of goods avoids delays caused by out-of-stock parts or inventory-carrying costs. 
Likewise, it can increase accountability, enhance mission reliability, reduce losses and theft of military equipment, and help with the time criticality on the military maintenance.

At the soldier level, tracking is useful in order to follow a proactive approach to logistics or to meet operational requirements. Soldier material (e.g., water, food, batteries or bullets) can be monitored with alerts issued for a necessary resupply. Aggregate data (e.g., groups of soldiers, companies, battalions...) might also be studied for further enhancements of supply for tactical and emergency units. The analytics might be focused on considering environment, body type, consumption...among other variables.

\subsection{Smart Cities Operations}
In denied area environments, existing IoT infrastructures could be reused in military operations. Ambient sensors can be used to monitor the existence of dangerous chemicals. Sensors monitoring human behavior may be used to assess the presence of people acting in a suspicious way. 
Leveraging information provided by pre-existing infrastructures might be critical.
Several security issues may arise, such as equipment sabotage or deceptive information. The authors of \cite{SecuredCity} categorize such attacks into four areas: (1) system architecture, firewalls, software patches; (2) malware, security policies and human factors; (3) third-party chains and insider threat; and (4) database schemas and encryption~technologies.

\subsection{Personal Sensing, Soldier Healthcare and Workforce Training}
Body-worn devices are increasingly available \cite{WearableComputing, SubcutaneousBAN}. Fitness trackers \cite{FitnessTrackers}
enable monitoring of physical activity along with vital signs.
This information has an obvious value for the users, but~there is also a significant potential in examining aggregate values of communities. Body-worn sensors, when~deployed on a community scale, offer information to support C4ISR.
We have to distinguish between participatory and opportunistic sensing. The last one may be of particular relevance for under-cover personnel involved in reconnaissance missions in urban environments.
Technologies for monitoring both workforce and their surroundings could aid when inferring physical or psychological states as well as assessing the risk of internal injury based on prior trauma \cite{SoldierFuture}. Soldiers can be alerted of abnormal states such as dehydration, sleep deprivation, elevated heart rate or low blood sugar and, if necessary, warn a medical response team in a base hospital. These wide range of health and security monitoring systems, enables an effective end-to-end soldier health system, including re-provisioning of health services when needed. 

In addition, IoT can be used in some training and simulation exercises, i.e., wearable receivers to mimic live combat. An example of live training may use cameras, motion and acoustic sensors to track force during training exercises. The system would send data to trainers' mobile devices, who can coach in real time and produce edited video and statistics to review after the exercise.

Other examples are Cubic's I-MILES (Instrumented-multiple Integrated Laser Engagement System) training solutions \cite{IMILES} which simulate combat using lasers and visual augmentation. They~use connectivity, computer modeling and neuroscience-based learning tools to provide a more comprehensive real-time training experience. The solutions simulate artillery fire and provide a battle effect simulator, which include explosive devices like land mines, booby traps, and pyrotechnics.

\begin{figure}[H]
\centering
\includegraphics[width=1\columnwidth]{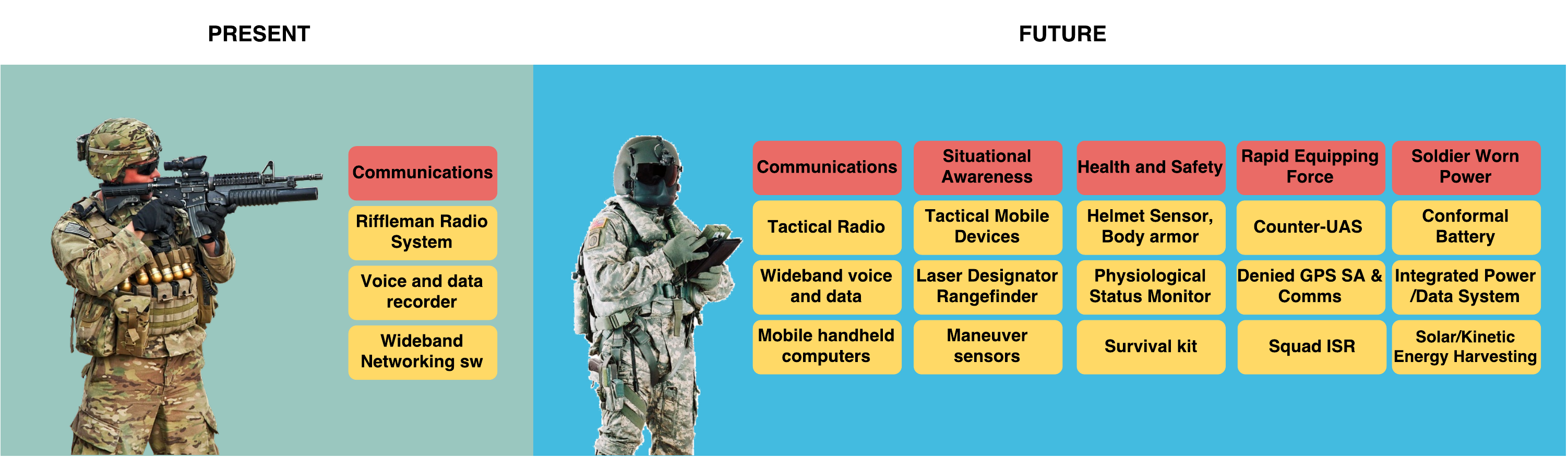}
\caption{Soldiers of today and the future.}
\label{fig:PersonVsSoldiersTodayAndTomorrow}
\end{figure}

The previously referred applications, and others yet-to-be imagined, could be part of the equipment of the soldiers of the future. A likely evolution of such equipment can be seen in Figure \ref{fig:PersonVsSoldiersTodayAndTomorrow}.

\subsection{Collaborative and Crowd Sensing}
Collaborative sensing involves sharing sensing data among mobile devices combined with robust short range communications. IoT nodes would be able to utilize placement or other sensors to supplement their own sensing methods. Once security issues (such as trust and authentication) are resolved, the information can be made available to the users. Long-term maintenance of IoT services yield multiple benefits, such as trend or fault detection. Individual sensor parameters must be considered to assign a particular relevance to a given reporting device and its feedback can be improved upon data fusion approaches.

IoT can ease ad-hoc mission-focused Intelligence, Surveillance and Reconnaissance (ISR) via pairing sensors with mission assignments \cite{SensorMission}. Thus, sensors and sensor platforms would not have to be burdened with excessive equipment to handle mission scenarios on their own. For example, multiple devices entering an area of interest, each with their own mission but relying on collaborative sensing to accommodate new or unanticipated requirements. In the case of a soldier, its situational awareness increases, allowing for improved survival and mission success.

Resource-rich devices might collect data from several sources to form a \ac{COP}.  This would allow for storing much of the collection and for processing the data locally.
Higher level functions would aid reducing response times, improving decision-making and reducing backhaul communications~requirements.

Crowdsensing promises to be an inexpensive tool for flexible real time monitoring of large areas and assessment for mission impact, hence complementing services potentially available in smart cities. Researchers have devised services that allow for decoupling applications from the hardware \cite{SensingService}. Crowdsensing services might offer both recruitment (devices that should be instructed to perform sensing) and data collections (the system returns data directly). These applications entail challenges. First, there is a large number of users that have to served while offering them tangible benefits or incentives \cite{incentives}.
Secondly, they are required to ensure a desired level of quality of the data collected.  

From the perspective of gathering data from a community, deception could be achieved by compromising individual devices. Consequently, the security level is proportionate to the number of present devices, each representing a possible attack vector. Moreover, the paradigm "Bring your own Device" (BYOD) \cite{BYOD} introduces potential security concerns, since the user may have full access and make use of multiple heterogeneous devices that are difficult to control.

Data validation is another domain-dependent task in the context of crowd sending, and, likewise, it is further complicated by the high heterogeneity of the devices. 
Even individual devices may vary in capabilities and performance over time. For example, for a smart phone, low battery levels could lead to infrequent GPS position updates, thus resulting in geo-tagging errors.
Approaches to ensure data quality often rely on oversampling and filtering of outlier values \cite{Zhang}, reputation schemes that provide trustworthy sensing for public safety \cite{Kantarci}, or trust network-based human intervention approaches.

Moreover, potential threats to privacy can compromise crowdsensing services. 
Many of these activities require geo-tagging and timestamps, which can lead to disclosing the locations visited by the users. Another privacy issue is the metadata collected about devices performing a sensing activity.

\subsection{Energy Management}
The U.S. DoD is already reducing its demand on facility energy by investing in efficiency projects on its installations \cite{DoDReport}. The introduction of data and predictive algorithms can help to understand usage patterns better and significantly decrease military's energy costs. 

\subsection{Surveillance}
Security cameras and sensors, combined with sophisticated image analysis and pattern recognition software, ease remote facility monitoring for security threats. In the case of marine and coastal surveillance, using different kinds of sensors integrated in planes, unmanned aerial vehicles, satellites and ships, make possible to control the maritime activities and traffic in large areas, keep track of fishing boats, and supervise environmental conditions and dangerous oil cargos.

Other examples can be the monitoring of hazardous situations: combustion gases and preemptive fire conditions to define alert zones, monitoring of soil moisture, vibrations and earth density measurements to detect dangerous patterns in land conditions or earthquakes, or~distributed measurement of radiation levels in the surroundings of nuclear power stations to generate leakage~alerts.

\section {Operational Requirements} \label{Requirements}

\begin{figure}[H]
\centering
\includegraphics[scale=0.85]
{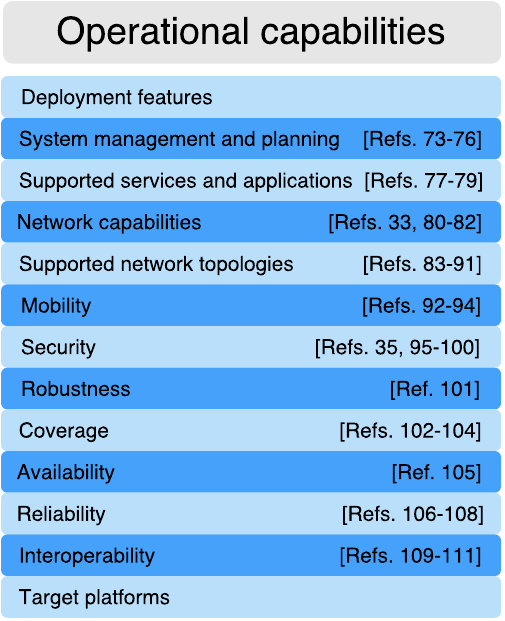}
\caption{Operational Capabilities assessed to cover mission-critical scenarios.}
\label{fig:OperationalCapabilities}
\end{figure}

The military has unique operational requirements \cite{ICMCIS}. Security, safety, robustness, interoperability challenges, as well as bureaucratic and cultural barriers, stand in the way of the broad adoption of new IoT applications.
In this Section a set of operational requirements grouped by capabilities (represented in Figure \ref{fig:OperationalCapabilities}) are assessed in order to cover the scenarios previously discussed.

\subsection{Deployment Features}
One of the biggest constraints in a battlefield environment is power consumption. IoT devices are likely to be powered by batteries or solar power, and charged on-the-move from solar panels, trucks, or~even by motion while walking. 
In either case, they should last for extended periods of time (at~least for the duration of the mission). Therefore, devices and sensors need to be power-efficient, and~end-users have to use them appropriately. Likewise, it is not easy to recharge IoT devices periodically or swap out batteries in deployed devices. Even in the case of body-worn devices, it is impractical to expect soldiers to carry additional batteries on top of their current equipment.

The exploitation of emerging embedded hardware within the military, probably through specialized software components designed to run on those innovative platforms, could lead to a significant increase in processing power and a decrease in energy consumption.

Furthermore, design values (e.g., power cell size or transmission capabilities) and equipment should fulfill the requirements imposed and be compliant with the considerations from military standards (e.g., MIL-STD 810G, MIL-STD 461F, MIL-STD-1275). IoT devices should be ruggedized and prepared to operate under extreme environmental conditions. Nevertheless, a non-negligible share of devices is already designed for harsh industrial environments, and, thus, they would be relatively well suited for the adoption in defense environments.

\subsection{System Management and Planning}
One of the largest gaps in the defense and public safety data ecosystem is digital analytics (data~collection, transformation, evaluation and sharing). Much of the massive information collected by sensors is never used, and, as for the information that is used, it often depends on manual entry and processing, which incur in significant delays when getting important information in mission-critical scenarios. Those delays can cause missions fail or stall, or force decision-making without relevant facts.
For example, USTRANSMCOM bulk supplies are tracked between major hubs using RFID tags, but~when supplies are broken down and distributed beyond the central hubs, they are replaced manually. Officers in the field sign for orders on paper and enter serial numbers into computers by hand. This approach is burdensome and poses risks due to human errors.

Much of the value of IoT is generated by automation, allowing systems to react quicker and with more precision than humans. Few military systems include fully autonomous responses.
For example, most unmanned systems deployed are not autonomous but remotely controlled by operators.
 
This management effort needs the development of new lightweight management protocols.
 For example, monitoring the M2M communications of IoT objects is important to ensure constant connectivity.
LightweightM2M \cite{LWM2M} is a standard developed by the Open Mobile Alliance (OMA) to interface between M2M devices and servers to build an application-agnostic scheme for the remote management of a variety of devices. The NETCONF Light protocol \cite{NETCONF} is an \ac{IETF} effort for the management of resource-constrained devices. In \cite{VanDenAbeele}, the authors propose a framework for IoT management based on the concept of intercepting intermediary nodes in which they execute heavy device management tasks on the edge routers or gateways of constrained networks.
The OMA Device Management working group specifies protocols for the management of mobile devices in resource constrained environments  \cite{OMADeviceMgmt}.

\subsection{Supported Services and Applications}
 
\begin{figure}[H]
\centering
\includegraphics[width=0.7\columnwidth]{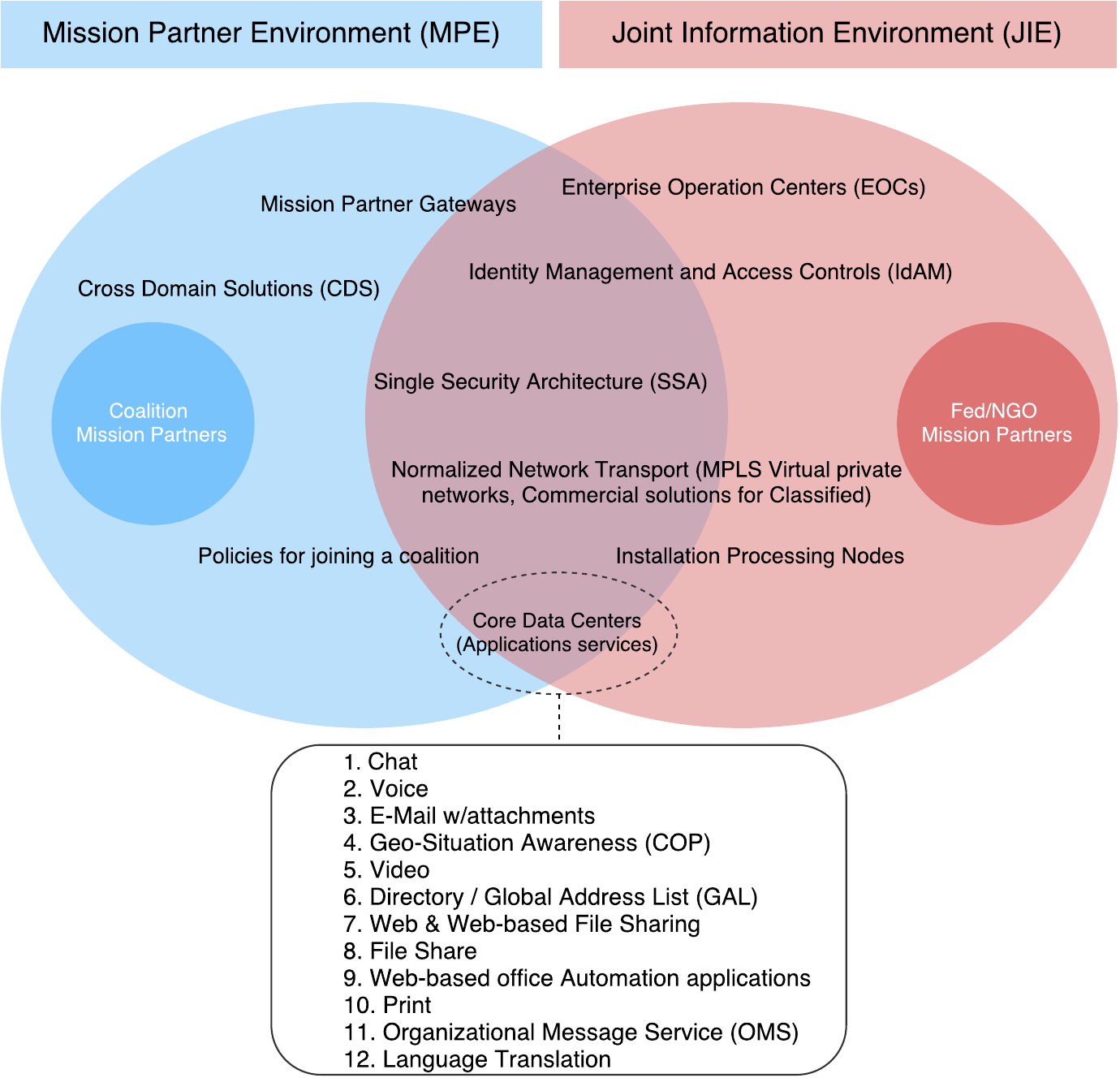}
\caption{Requirements and application services for commanders.}
\label{fig:Services}
\end{figure}

A number of commercial devices and electronic equipment have been explored to provide the services required, like chat, push-to-talk voice, geo-situational awareness, SRTV (Secure Real-Time Video) or web sharing. A complete list of requirements and application services can be seen in Figure~\ref{fig:Services}. The diagram represents the vision of the \ac{JIE}, which ensures that DoD military commanders, civilian leadership, warfighters, coalition partners, and other non-DoD mission partners, access information and data provided in an agile DoD-wide information environment. This shared IT infrastructure includes enterprise services, and a Single Security Architecture (SSA).
The \ac{MPE} is integrated with and enabled by JIE. It corresponds to an operating environment that enables C2, within an specific coalition, for operational support planning and execution on a network infrastructure at a single security level with a common language. Regarding the small circles of the diagram, they represent the different participants within a specific partnership or coalition (e.g., the intelligence community, U.S. government agencies, allies, and other mission partners, such as industry organizations and Non-Governmental Organizations (NGOs)).
 
\begin{figure}[H]
\centering
\includegraphics[width=1\columnwidth]{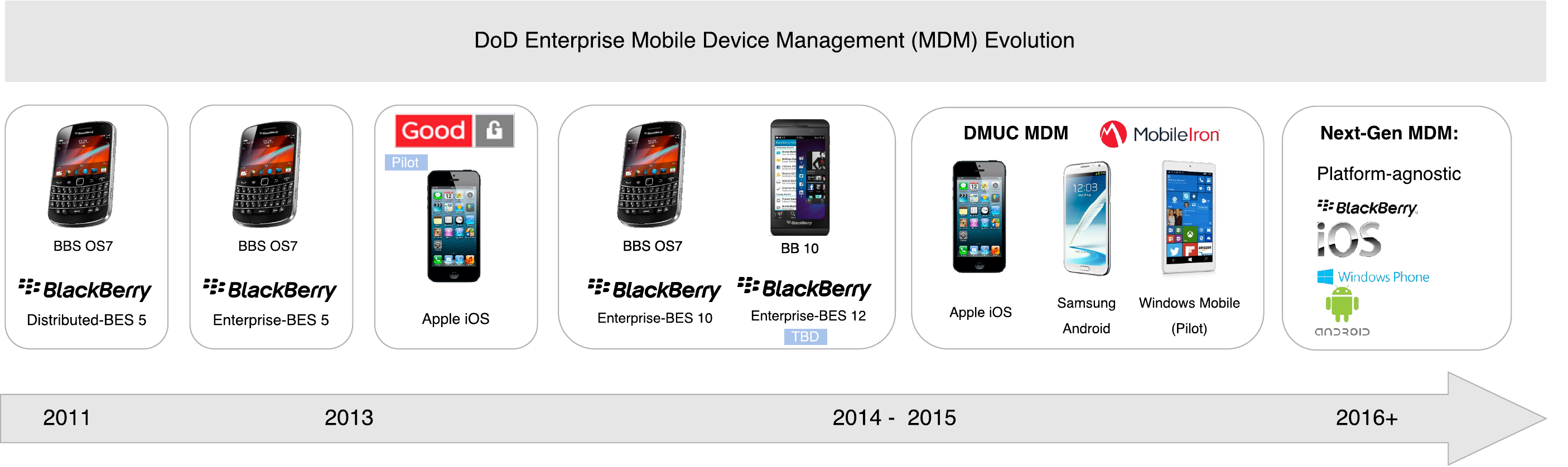}
\caption{DoD
%Please define.
 Enterprise Mobile Devices Management (MDM) evolution.}
\label{fig:Devices}
\end{figure}

The U.S. army's Nett Warrior (NW) program \cite{Portfolio2014} has developed ruggedized Android devices. These devices, which are modified from COTS Samsung Galaxy Note II smart phones, provide access to the data-capable Rifleman radio. It aims to connect soldiers in the field with a range of apps, such as Blue Force Tracking, 3-D maps, or an application that shows details on profiles of high-value targets. The devices run an NSA-approved version of the Android operating system, and plan to include applications such as foreign language translation. These programs have been piloted on a limited basis. Broader deployment is hampered by the limited usability, functionality and lack of connectivity. Other commercial devices can be seen in Figure \ref{fig:Devices}.

The U.S. Air Force (Hurlburt Field, Florida, United States) has developed apps on commercial iPads.  For example, programmers at Scott Air Force Bases created in 2014 an app to plan loads for the KC-10 cargo aircraft \cite{TabletApp}, winning an award for government innovation. 
Such an application was designed to automatically gauge pre-flight distribution of cargo in a weight and balance computation, considering the crew, fuel and cargo in a drag-and-drop interface.

\begin{figure}[H]
\centering
\includegraphics[width=0.9\columnwidth]{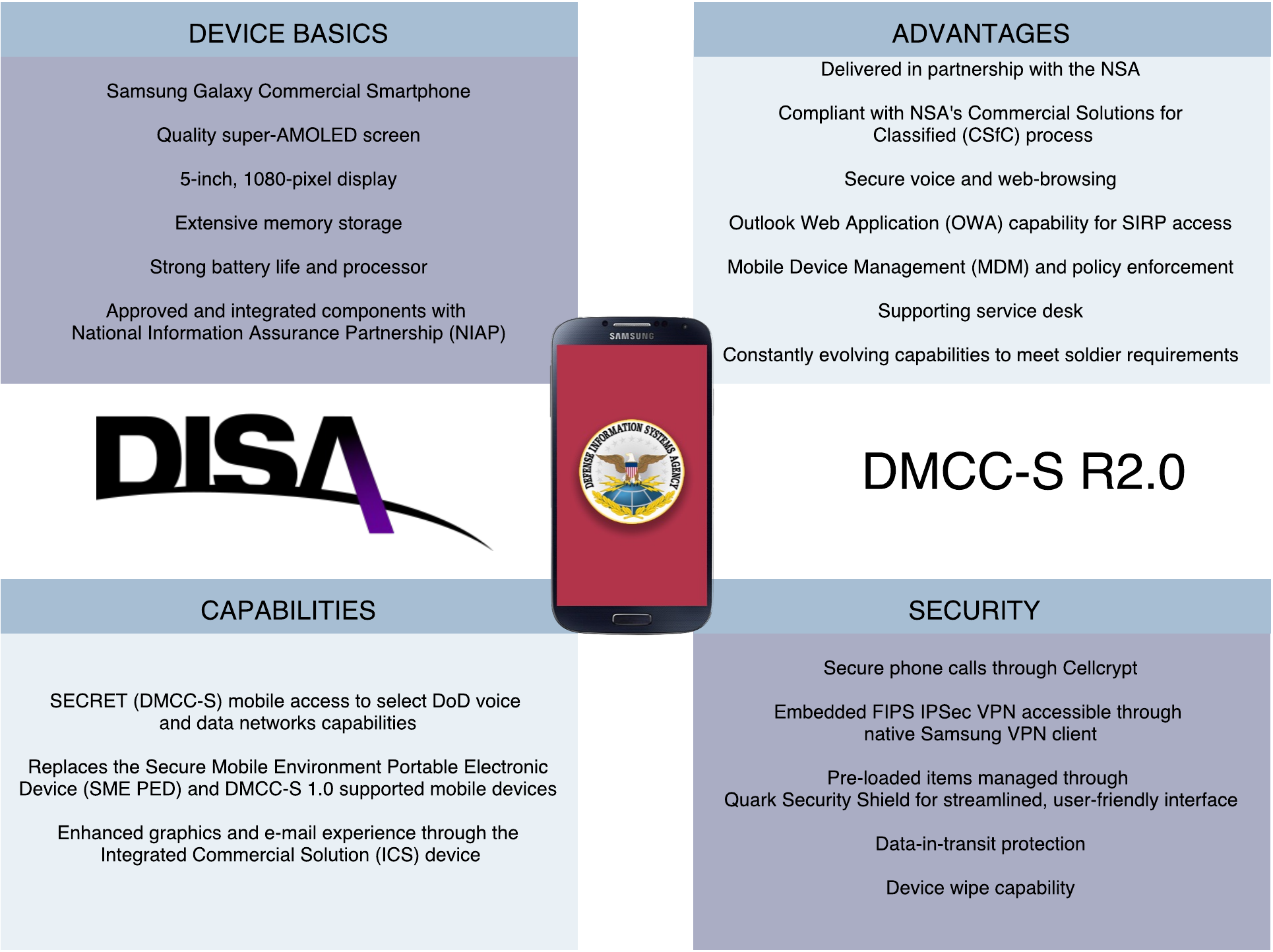}
\caption{Main characteristics of DMCC-S R2.0}
%Please define.
\label{fig:DISA}
\end{figure}

The American Defense Information System Agency's (DISA) Mobility Program has implemented software packages for NSA-approved Android devices. The program includes secure devices that can access a secret classified network, SIPRNET \cite{NCW}. DMCC-S (DoD Mobility Classified Capability Secret Device) R2.0 is an example (Figure \ref{fig:DISA}) of the new generation of DoD secure mobile communication~devices.

\subsection{Network Capabilities}
Military network infrastructures are severely limited by frequent disconnections, partitioning, and~fluctuations of radio channel conditions. This can lead to issues in sensing availability and constraints on the usage of transducers. The military IoT networks operate over tactical radios that establish and maintain mobile and fixed seamless C2 communications between operational elements and higher echelon headquarters. Tactical radios provide interoperability with all services, various agencies of the U.S. Government, commercial agencies, and allied coalition forces. High-bandwidth radios that could constitute integrated networks are still under development. For example, Harris~Corporation  (Melbourne, Florida, United States) will deliver the first batch of RF-335M tactical radio systems in September 2017 to U.S. Special Operations Command (USSOCOM) \cite{Harris}.

As the connectivity of sensors improves, the system can become overwhelmed with the huge volume of data in transit.  This increase in volume of data may force an upgrade to a system's network infrastructure to increase bandwidth or, alternatively, it may require to increase the performance of intelligent data filtering.
In the commercial environment, network bandwidth and QoS (Quality~of Service) challenges are addressed using COTS hardware combined with open virtualization platforms to manage network demands dynamically. These advanced network servers provide both high availability and also new approaches to control and provision network systems by delivering a path to Network Function Virtualization (NFV) \cite{NFV}. NFV offers the operator the ability to configure the network infrastructure  dynamically through sophisticated management protocols. Thus, NFV~empowers military commanders to quickly configure data feeds for changing operational requirements, and to manage device and data security throughout the system.
For example, the~use of OpenStack~\cite{Openstack} optimizes different network demands, such as giving priority to certain data flows, or~protecting parts of the network from attacks. Commercial solutions can be adapted to the management of data as  they pass through various military networks to get to the combat cloud.

Currently, each military force has its own infrastructure, both for connectivity and for the back-office systems.
Transitioning to a combat cloud infrastructure will offer greater ability to export both assets and data in the field for joint operations.
Also a combat cloud will allow information and control to move forward when appropriate, providing the operational flexibility to deal with coalitions.

Nowadays, any army in the world can have the network infrastructure needed to handle, process and distribute the massive flow of data that would be generated by a widespread IoT \cite{Deloitte}. In order to make effective use of IoT, the devices must be able to connect to global networks to transmit sensor data and receive actionable analytics. 

\subsection{Supported Network Topologies}

Mobile ad-hoc networks (MANET) \cite{Manet,Manet2} and hybrid wireless sensor networks (WSNs) \cite{WSN, WSN2} are the main tactical topologies.The authors of \cite{NetworkCoding} evaluate the performance of network coding in the context of multi-hop military wireless networks. The researchers prove its efficiency in multicast and broadcast communications. They also test the optimal capacity of the system and its ability to recover from lost packets.
Network coding operates well even with highly lossy and unreliable links.

Interest has grown in opportunistic sensing systems \cite{OS}, particularly on those that take advantage of smartphone-embedded sensors \cite{Quake}. 
A network of opportunistic sensing systems can automatically discover and select sensor platforms based on the operational scenario, detecting the appropriate set of features and optimal means for data collection, obtaining missing information by querying resources available, and using appropriate methods to fuse data. Thus, the system results in an adaptive network that automatically finds scenario-dependent objective-driven opportunities with optimized performance. For example, Mission-Driven Tasking of Information Producers (MTIP) \cite{MTIP}, is a prototype system for sharing of airborne sensors that focuses on the effective allocation of a large number of potentially competing individual tasks to individual sensors. Nevertheless, specific protocols are needed for advancing autonomous sensing that not only ensures effective utilization of sensing assets but also provides robust optimal performance \cite{futureOS}.

Moreover, the development of a decentralized infrastructure is needed to avoid a single point of failure. Bandwidth is perhaps one of the most precious resources in a tactical environment.
It~is expected that in dynamic battlefield environments large-scale data analysis will be conducted in near real-time. This fact implies constraints on data analysis coupled with connectivity challenges.
Decentralizing computational resources by creating multiple and local cloudlets is insufficient if the overall approach still consists in sending raw data from transducers to a local cloud for processing. 

\subsection{Mobility Capabilities}
Mobility is another challenge for the IoT implementations because most of the services are expected to be delivered to users on the move.
Service interruption can occur when the devices transfer data from one gateway to another. 
To support service continuity, Ganz et al. \cite{Ganz} propose a resource mobility scheme that supports two modes: caching and tunneling. These methods allow applications to access IoT data when resources become temporarily unavailable. The evaluation results show a reduction of service loss in mobility scenarios of 30\%.

\begin{figure}[H]
\centering
\includegraphics[width=1\columnwidth]{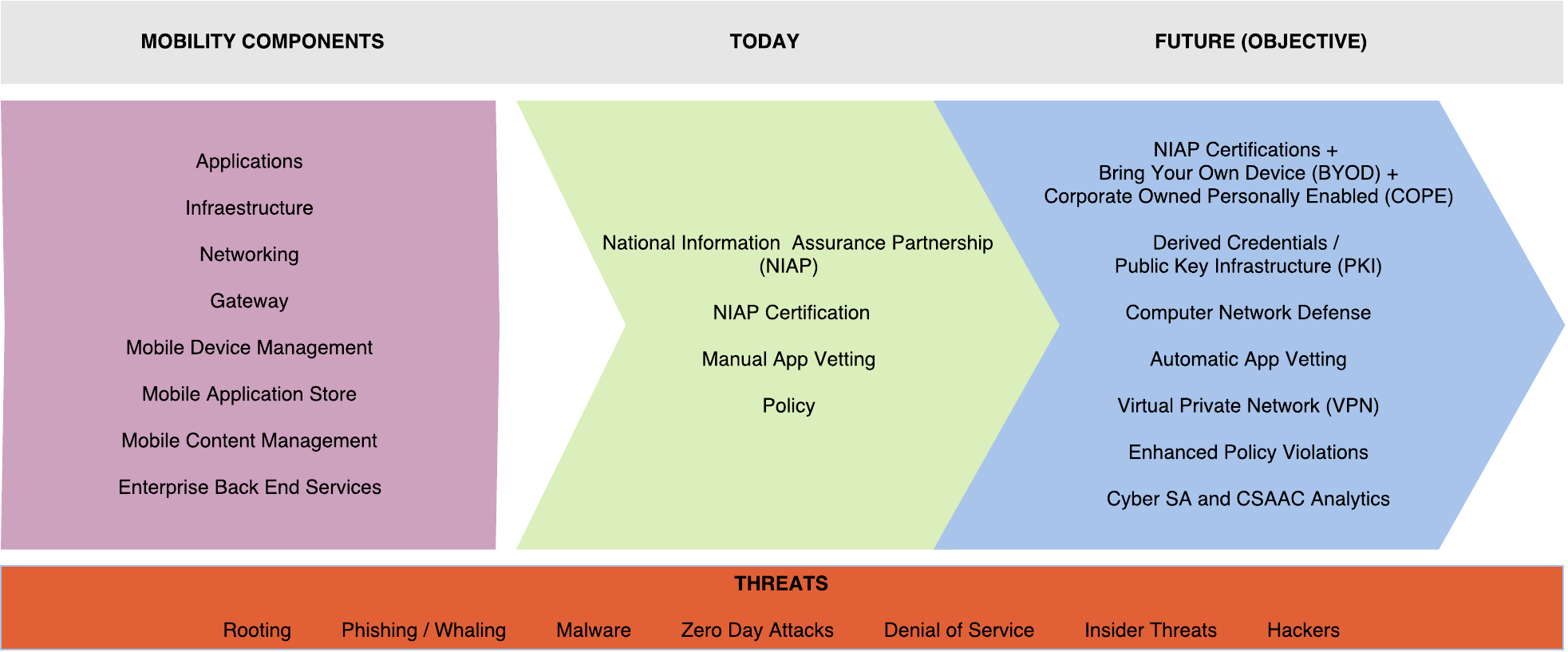}
\caption{Mobility components and their security.}
\label{fig:Security}
\end{figure}

The huge number of smart devices in IoT systems also requires efficient mechanisms for mobility management (the components and their needs are illustrated in Figure \ref{fig:Security}).
For instance, a feasible approach for M2M communications is presented in \cite{Fu}. In the scheme presented, group mobility is managed by a leader based on the similarity of their mobility patterns.

Various solutions support mobility for vehicle-to-vehicle networking (Internet of Vehicles (IoV) solutions).
An example of a group mobility mechanism for MANET is presented in \cite{Misra}, where the authors were inspired by the mobility of a flock of birds flying in a formation.

\subsection{Security Capabilities}

Security is a paramount challenge that needs to be addressed at every level of IoT, from the high volume of endpoint devices that gather data and execute tasks, to cloud-based control systems through network infrastructure \cite{Wrona}.  
Several protection mechanisms and Electronic Protection Measures (EPM) features have been identified within
TRANSEC such as Low Probability of Interception (LPI) (e.g., secure PN-sequence generators to prevent easy sequence estimation, scrambling of transmission data and control information),  Low Probability of Detection (LPD) (e.g., frequency hopping and spread-spectrum techniques) and Anti-Jamming (AJ), INFOSEC (e.g., NATO Level 3 security including IP security protocols (IPSec/HAIPE) as well as IP
tunneling protocols), COMSEC and NETSEC~capabilities.

Privacy issues and profile access operations between IoT devices without interference are critical. 
IoT nodes require a variety of widely-used and well-established security mechanisms (e.g., SSL, IPSec, PKI) that perform numerous computationally intensive
cryptographic operations \cite{ECC}. The use of smaller cryptographic key sizes is examined in \cite{SmallKeys}. 
Sensitive data needs a transparent and easy access control management.
Several proposals can be used, such as grouping devices, or present only the desired devices within each virtual network.
Another approach is to support access control in the application layer on a per-vendor basis. 
Relevant projects that estimate the network location of objects to perform context-aware services are reviewed in \cite{ContextAwareness}. Current methods for location estimation are based on IP. However, Named Data Networking (NDN) is one of the candidates for naming infrastructure in the future Internet \cite{NDN}.

Military equipment can be subject to either interference, sabotage, potential manipulation or disruption of data flows between different units, resulting either in service interruptions, intrusions,  propagation of misinformation, or misleading the COP on the needs of support units. These failures in equipment can compromise both intelligent gathering and planned operations having obvious mission and life-threatening consequences. 

For example, inadequately secured networks can provide the enemy with intelligence (location, deployment), allowing the adversary to anticipate movements of forces.

Furthermore, security vulnerabilities could allow enemies to take control or disable automated systems, preventing workforce from carrying out their mission, or even using their own assets against them. Next, we describe the main security challenges:

\begin{itemize}[leftmargin=*,labelsep=5mm]
\item Device and network security: the potential of IoT is derived to a large extent from the ubiquity of devices and applications, and the connections between them. This myriad of links creates a massive number of potential entry points for cyber-attackers. The systems also depend on backbone storage and processing functions, which can include other potential vulnerabilities. One~of the ways to enhance the security of a complex network is to limit the number of nodes that an attacker can access from any given entry point. This approach conflicts with IoT, which~generates much of its value from the integration of different systems. Securing a broad range of devices is also difficult. Many of them have limited capacity with no human interface and depend on real-time integration of data. This complicates traditional approaches to security, like multi-factor authentication or advanced encryption, which can hinder the exchange of data on the network, requiring more computing power on devices, or needing human interaction.

\item Insider misuse: cyber risks and insider threats are a challenge for large organizations. A single mistake from a single user can allow an attacker to gain access to the system.

\item Electronic warfare: most technologies communicate wirelessly on radio frequencies. Adversaries can use jamming techniques to block those signals making the devices unable to communicate with backbone infrastructure. Wireless connections also raise the risk of exposing the location through radio frequency emissions. Transmitters can serve as a beacon detectable by any radio receiver within range, and the triangulation of such emissions can compromise the mission.

\item Automation: the full automation of equipment and vehicles extends the reach of cyber threats to the physical domain.

\end{itemize}

The authors of \cite{IntegrityAttestation} propose integrity attestation as a useful complement to subject authentication.
Thus, the provision of a data structure can convey integrity assurances and be validated by others.
This~is particularly useful for IoT, considering that the limited capacity of the computers
and communication channels do not allow for complex protocols to detect malfunctions. The~document~\cite{CloudSecRequirements} outlines the DoD security model to leverage cloud computing along with the security requirements needed for using commercial cloud-based solutions.

\subsection{Robustness Capabilities}
Communication technologies will provide robustness to signal interference and/or loss of network operation. When deployed in locations with other tactical networks (i.e., vehicular deployment), proper measures to avoid interference from adjacent users will be needed. For mesh or  Point-to-Multi-Point (PMP) modes, the network will provide redundancy and be robust to a single point of failure. Systems should be robust to jamming, supporting techniques to actively track jamming signals and applying automatic jamming avoidance measures. It should include cognitive radio and dynamic spectrum management
techniques to automatically overcome bad conditions in the communications~environment.

The operational requirements for robustness also include the physical attributes of the device. Generally, this is addressed by the target platform requirements, which, in turn, is dependent on the deployment scenario. Equipment should be also physically robust to environmental damage, i.e., shock-~and water-proof. The IoT system should provide mechanisms to allow for fast switching between the technology chosen and back-up/legacy communications in the event of failure. Although there are many metrics available to assess the performance of IoT devices, evaluating their performance is a challenge since it depends on many components as well as the behavior of the underlying technologies. The evaluation of routing protocols, information processing  \cite{GanzEvaluation}, application layer protocols, and QoS have been reported in literature, but there is a lack of a thorough performance evaluation for IoT~services.

\subsection{Coverage Capabilities}
Defense and PS should invest in resilient, flexible and interoperable capabilities to operate at extended ranges under adverse weather conditions and harsh environments (including \ac{LOS} and \ac{NLOS} scenarios) in enemy territory, and enhance connectivity in denied areas. One of the technologies that can deliver mobile and persistent connectivity is CubeSat: nano satellites that can be deployed in large number to create potentially more resilient constellations~\cite{CubeSat}. 
CubeSat deployment is also faster than in larger satellites, as they can be launched into orbit in clusters or piggybacked on other loads. 
It supports \ac{SDR} to enable reconfigurability of data management, protocols, waveforms, and data protection. 

Other technologies are High-Altitude Platforms (HAPs) and Unmanned Air Vehicles (UAVs), such~as drones that operate above the range of terrestrial communication systems and can be equipped with communication relays.
Unlike satellites, which eventually become defunct, HAPs can be upgraded and enhanced as technologies evolve. They also have significant advantages over manned communications platforms, as they can stay airborne continuously for long periods.

The U.S. military has already deployed four EQ-4B Global Hawk Block 20 Drones with the Battlefield Airborne Communications Node (BACN) system, but it will need significantly greater capacity to deliver connectivity to a full suite of connected devices across multiple theaters.
DoD~is now involved in the development of Northrop Grumman RQ-4 Global Hawk Block 30 and 40, ground~stations, and Multi-Platform Radar Technology Insertion programs. The U.S. Navy will get a persistent maritime ISR capability through the MQ-4C Triton. DoD is now funding the procurement of two Low Rate Initial Production (LRIP) systems and continues to
fund development activities associated with software upgrades \cite{UAV, UAVRoadmap}.

\subsection{Availability}
Availability must be taken into account in the hardware, with the existence of devices compatible with IoT functionalities and protocols; and in the software, with available services for everyone at different places.
One solution to achieve high availability is to provide redundancy for critical devices. Moreover, there are studies on assessing the availability of IoT applications at the first design stages~\cite{Macedo}.

\subsection{Reliability}
The critical part to increase the success rate of IoT service delivery is the communication network. 
The authors of \cite{Maalel} propose a reliability scheme at the transmission level to minimize packet losses. Other authors \cite{Li} exploit probabilistic model methods to evaluate the reliability and cost-related properties of the service composition in IoT systems. 
The survey \cite{MarkovWSN} reviews  applications
of the Markov decision process (MDP) framework, a powerful decision-making tool to develop adaptive algorithms and protocols for WSNs, like data exchange and topology formation, resource and power optimization, area coverage, event tracking solutions,
and security and intrusion detection methods.

\subsection{Interoperability Capabilities}

Taking advantage of the full value of IoT is about maximizing the number of hardware and software systems, nodes and connections in the data ecosystem.
However, defense lacks a cohesive IT architecture. The different and heterogeneous systems are developed independently and according to different operational and technical requirements. 
Frequently, multiple services are involved in an operation, or several departments are involved in a process, but information has to be adapted between their systems manually. The usage of different hardware designs and data standards can impact the cohesion of defense infrastructure, leading to stove pipe systems. 
The fragmentation of the architecture also complicates the use and development of common security protocols. Adequate interoperability between devices is often not achieved given the variety of functions served by defense hardware, the integration across partners, or when potentially useful devices in an area of operations are to be leveraged (i.e., smart city deployment). IoT capabilities across an enterprise as broad as defense can only be delivered through a suite of common standards and protocols. 

To enhance end-to-end interoperability, one of the most popular approaches is the usage of Service-oriented Architectures (SoAs). 
SoAs use common messaging protocols and well-defined interfaces to share information between multiple services.  They consider aspects such as service reuse, rapid configuration, and composability with dynamic workflows. 
SoAs in the tactical domain could help to leverage commercial IoT capabilities and attempt to address the interoperability challenges specific to \ac{C4ISR}. 

Both military computers and sensor networks should have longer service lives than commercial equivalents, resulting in greater needs to maintain legacy systems.
One of the key weaknesses of legacy systems is their lack of interoperability. This limits significantly the ability to integrate new platforms into the defense digital ecosystem, and to leverage existing systems in innovative ways.
 
DISA is implementing a cohesive digital architecture through the Joint Information Environment (JIE) initiative \cite{JIE}  to unify capabilities, facilitate collaborations with partners,  consolidate infrastructure, 
create a single security architecture, and provide global access to services.

TacNet Tactical radios \cite{LockheedMartin} help to demonstrate how an open systems architecture can enable improved interoperability between next-generation and legacy fighter aircrafts. Lockheed Martin performed tests on a F-22 and F-35 Cooperative Avionics Test Bed (CAT-B). Those aircrafts were flown to assess the capability to share real-time information among varied platforms. The ability to transmit/receive Link-16 communications on F-22 was proven, also the software reuse and reduction of the aircraft system integration and the use of Air Force UCI messaging standards. 

U.S. Army CERDEC NVESD \cite{ISA} has developed ISA under the Deployable Force Protection program. ISA is an interoperability solution that allows components to join a tactical network and use its functionality without requiring neither prior knowledge of the resources available on that network, nor physical integration.  ISA uses dynamic discovery to find other ISA-compliant systems, regardless of platform. This dynamic discovery is accomplished by requiring all members to announce the data they provide and functionality they can perform when they connect to the network. Members can change their capabilities on the fly and search for others that provide either data or the functionality they need. ISA understands the capabilities of those sensors and shares their information with operators. 
When future sensors come online to a network, they can register and communicate their capabilities. Assets and sensors on that network can then subscribe to the types of information they are interested in.
ISA seeks to provide the critical capabilities needed for a forward operating base to defend itself. It improves the mobile Soldier's situational awareness by enabling him to query different sensors as he moves through an area and access to information that was previously unseen to him, such as event messages. 

\subsection{Target Platforms}
The complexity and high cost of defense systems mean that they will remain in service for years. As we have previously indicated, the longevity of ground/airborne/seaborne platforms, and a form factor designed for handheld or manpack use, creates interoperability issues as well as operational challenges when enhancing their capabilities and attaching them to the combat cloud. New~technologies such as multi-core silicon and virtualization can create affordable solutions. 
On~legacy single-core processors this virtualization would have a direct impact on platform performance. The processor will have to run both legacy and new code while maintaining strict separation for safety and security reasons. 
With multi-core technology the performance and separation risks can be mitigated in silicon (with separated cores for legacy and new environments, and~separated~networks).

\section{Building IoT for Tactical and Emergency Environments}
\label{BasicsIoT}
\begin{figure}[H]
\centering
\includegraphics[width=0.9\columnwidth]{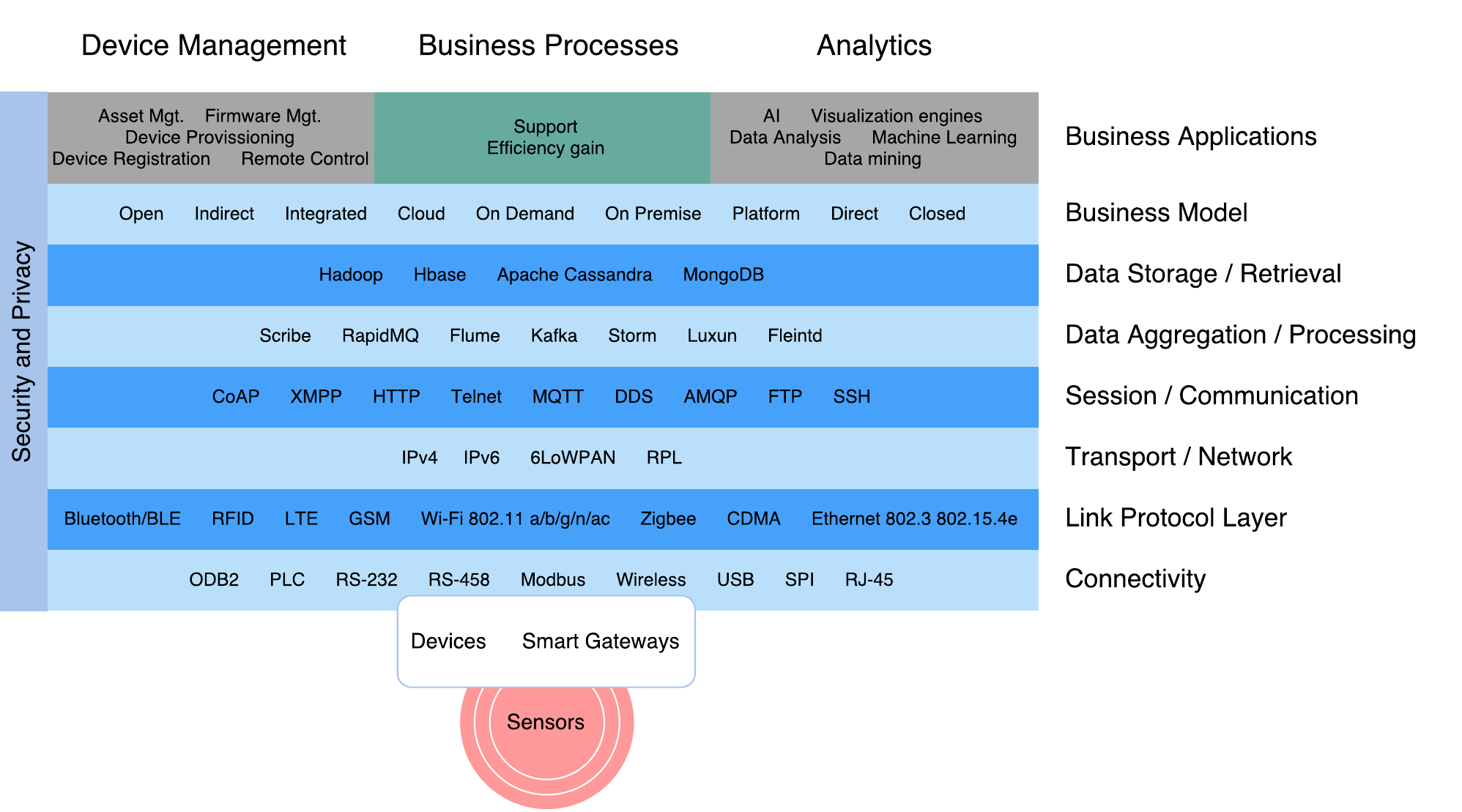}
\caption{IoT landscape.}
\label{fig:IoTLandscape}
\end{figure}

In order to understand the complex adoption of IoT for defense, this section will briefly review the basics of IoT landscape (a graphical overview of the main elements can be seen in Figure \ref{fig:IoTLandscape}) to support the requirements previously explained. First, it focuses on the architecture with an overview of the most important elements. Next, the section examines the main standardized protocols and~technologies.

\begin{figure}[H]
\centering
\includegraphics[width=1\columnwidth]{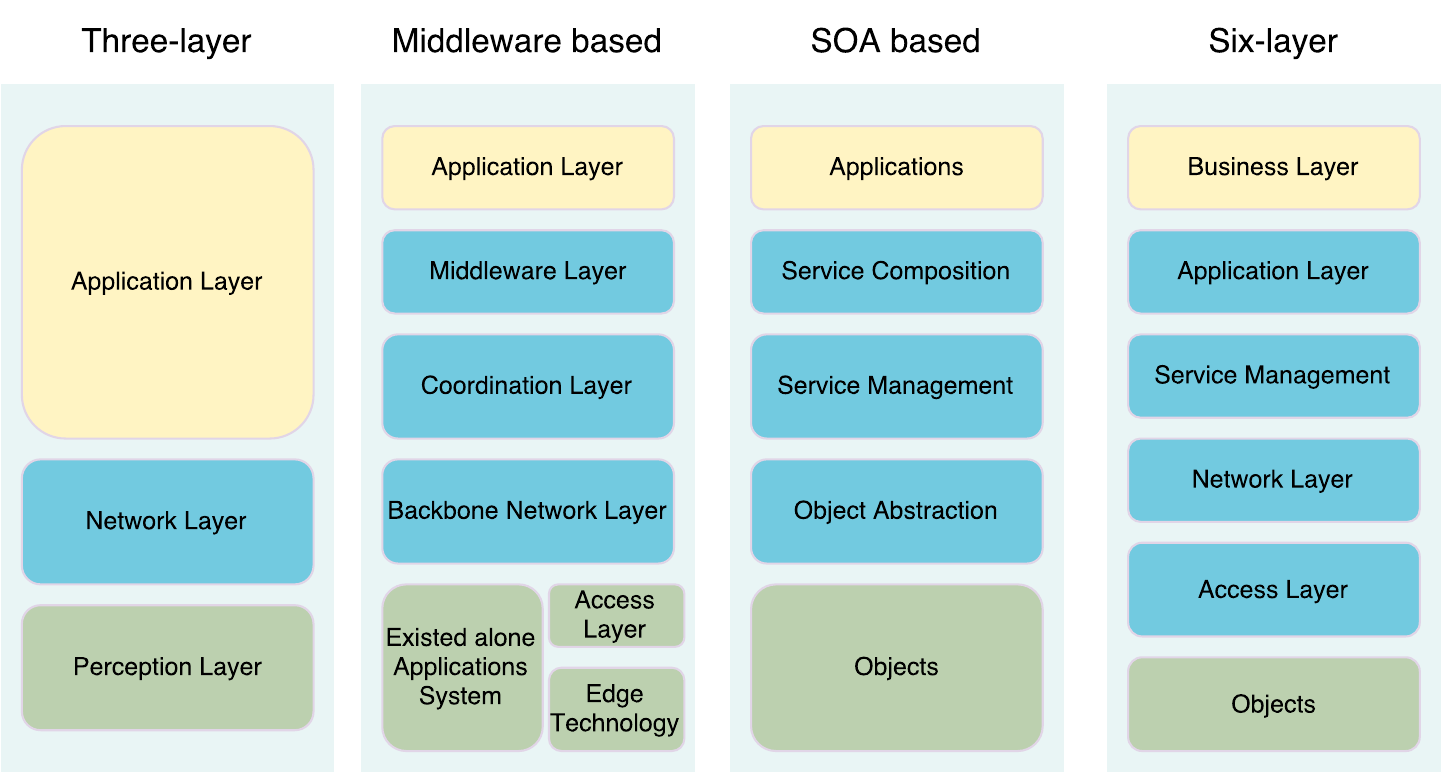}\\
\begin{tabular}{cccc}
(\textbf{a}) ~~~~~~~~~~~~~~~~~~~~~~~~~~~~~~~~~~~~~~~~~~~&(\textbf{b})~~~~~~~~~~~~~~~~~~~~~~~~~~~~~~~~~~~~~ &(\textbf{c})~~~~~~~~~~~~~~~~~~~~~~~~~~~~~~~~~~~~~~ &(\textbf{d})\\
\end{tabular}
\caption{The IoT architecture. ({\bf a}) Three-layer; ({\bf b}) Middleware-based; ({\bf c}) Service-Oriented Architecture (SOA)-based; ({\bf d}) Six-layer.}
\label{fig:ArchitectureModels}
\end{figure}

The increasing number of IoT proposed architectures has not converged to a reference model~\cite{Krco} or a common architecture~\cite{fp7}. 
In the latest literature, it can be distinguished among several models, as it can be seen in Figure \ref{fig:ArchitectureModels}. For example, the basic model which has three layers (application, network and perception layers) was designed to address specific types of communication channels and does not cover all the underlying technologies that transfer data to an IoT platform. 

Other proposals include a middleware based layer \cite{middleware}, a Service-Oriented Architecture (SOA) based model \cite{SoA}, and a six-layer model.
There are differences between these models: for example, although the architecture is simpler in the three-layer model,  layers are supposed to run on resource-constrained devices, while a layer like ``Service Composition'' in the SOA-based architecture takes a rather big fraction of the time and energy of the device.

Next, we provide a brief description on the functionality of the most common layers.

\begin{itemize}[leftmargin=*,labelsep=5mm]
\item{Perception layer}: this first layer represents the physical elements aimed at collecting and processing information.
Most COTS IoT devices are designed for benign environments and currently focus on home automation, personal services and multimedia content delivery. Miniaturized devices such as transducers (sensors and actuators), smartphones, System on Chips (SoCs) and embedded computers
are getting more powerful and energy efficient. The next generation of processors includes new hardware features aimed at providing highly trusted computing platforms. For~example, Intel includes an implementation of the Trusted Platform Module (TPM) designed to secure hardware through cryptography.
Technologies such as ARM TrustZone, Freescale Trust Architecture, and Intel Trusted Execution enable the integration of both software and hardware security features \cite{Wind}. Plug-and-play mechanisms are needed by this layer to configure heterogeneous networks. Big data processes are initiated at this perception layer. This~layer transfers data to the Object Abstraction layer through secure channels. 

\item{Object Abstraction Layer}: it transfers data to the Service Management layer through secure channels. 
To transfer the data, the protocols used in the COTS IoT nodes either use existing wireless standards, or an adaptation of previous wireless protocols in the target sector. Typically, IoT devices should operate using low power under lossy and noisy conditions. Other functions like cloud computing and data management processes are handled at this layer \cite{Yang}.

\item{Service Management Layer or Middleware}: this layer enables the abstraction of specific hardware platforms. It processes the data received, takes decisions and delivers the services over network protocols \cite{Chaqfeh}.

\item{Application Layer}: it provides the services requested to meet users' demands.

\item{Business Management Layer}: this layer designs, analyzes, develops and evaluates elements related to IoT systems, supporting decision-making processes based on Big Data.
The control mechanisms for accessing data in the Applications layer are also handled by this layer. It builds a business model based on the data received from the Application layer. 
Moreover, this layer monitors and manages the underlying four layers, comparing the output of each one with the output expected to enhance services and maintain users' privacy \cite{Wu}.
This layer is hosted on powerful devices due to its complexity and computational needs.
\end{itemize}

\begin{figure}[H]
\centering
\includegraphics[width=0.9\columnwidth]{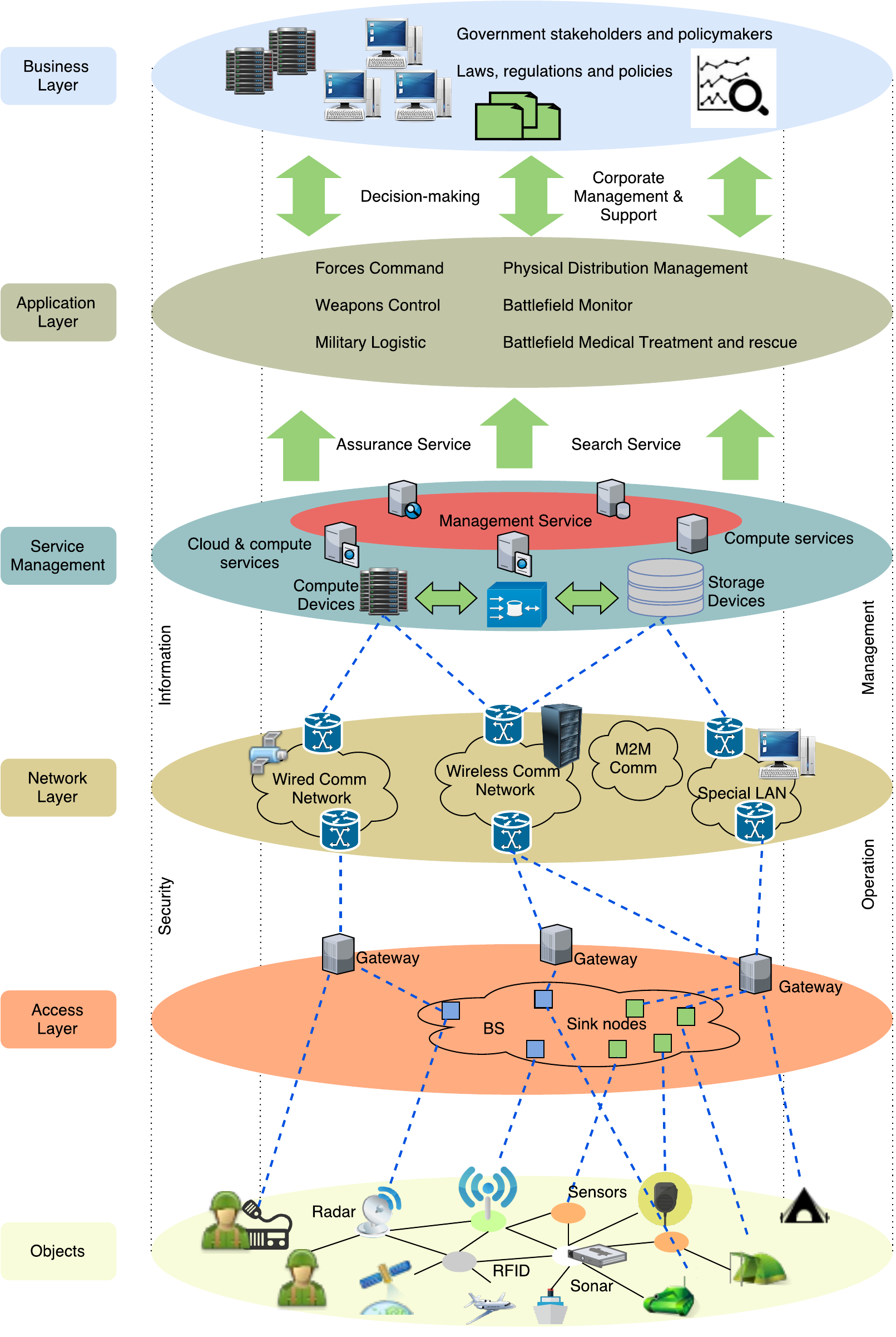}
\caption{Example of military architecture with six layers.}
\label{fig:MilitaryArchitecture}
\end{figure}

A generic IoT architecture is presented in \cite{Sarkar}. It introduces an IoT daemon consisting in three layers with automation, intelligence and zero-configuration: Virtual Object, Composite Virtual Object, and Service layer. An example of a possible military architecture can be seen in Figure \ref{fig:MilitaryArchitecture}.

The process of sensing consists in collecting data from objects within the network, and sending them back to a data warehouse, a database or a cloud system, to be analyzed and act. Four main classes of IoT services can be categorized:
\begin{itemize}[leftmargin=*,labelsep=5mm]
\item Identity-related services: these services are employed to identify objects,  but are also used in other types of services.
\item Information Aggregation services: these services collect and summarize raw measurements.
\item Collaborative-Aware services: these services act on top the Information Aggregation services and use the obtained data to make decisions.
\item Ubiquitous services: these collaborative-aware services function anytime to anyone, anywhere.
\end{itemize}

\afterpage{\clearpage}

Most existing applications provide the first three types of services.  The ultimate goal are the ubiquitous services.

Semantic analysis is performed after sensing to extract the corresponding knowledge. It includes discovering, resources usage and information modeling. Thereafter, recognizing and analyzing data to take proper decisions within the service. This is supported by semantic web technologies \cite{NATOSematic} such as the Resource Description Framework (RDF), the Web Ontology Language  (OWL), or Efficient XML Interchange (EXI), adopted as a W3C recommendation.

\subsection{IoT Standardized Protocols}

The U. S. Defense Standards, also called Military Standards (MIL-STD), are used to help achieve standardization objectives. These documents are also used by other non-defense government organizations, technical organizations, and industry. The ASSIST database \cite{MIL-STD} gathers these documents and also includes international standardization agreements, such as NATO standards, ratified by the United States and International Test Operating Procedures. Furthermore, the DoD is starting to use civilian standards, since numerous contributions to the deployment and standardization of the IoT paradigm come from the scientific community. Among them, the most relevant are the ones provided by the European Commission and the European Standards Organisations (i.e., ETSI, CEN, CENELEC), by their international counterparts (i.e., ISO, ITU), and by other standards bodies and consortia (W3C, \ac{IEEE}, EPCglobal). The M2M Workgroup of the \ac{ETSI} and some \ac{IETF} Working Groups are particularly important.
 
In this section we provide an overview of some of the standardized protocols that could be used for providing the IoT services described in the previous sections.

\subsubsection{Application Layer Protocols}
The following are the most popular Application Layer protocols: Constrained Application Protocol (CoAP), Message Queue Telemetry Transport (MQTT), Extensible Messaging and Presence Protocol (XMPP), Advanced Message Queuing Protocol (AMQP) and Data Distribution Service (DDS). Performance evaluations and comparisons between them have been reported in the literature \cite{comparison}. Each of these protocols may perform rather well in specific scenarios, but there is no evaluation of all these protocols together. Consequently, it is not possible to provide a single prescription for all IoT applications, just that they must be designed from the ground up to enable extensible operations \cite{Uckelmann}.

\subsubsection{Service Discovery Protocols}
 
Resource management mechanisms are able to register and discover resources and services in a self-configured, efficient and dynamic way. Such protocols include CoAP resource discovery, CoAP Resource Directory (RD), and DNS Service Discovery (DNS-SD)  which can be based on mDNS (Multicast DNS). A detail description of their characteristics can be seen in \cite{serviceDiscovery}.

\subsection{Enabling Technologies}

%Table \ref{TabProtocol} provides an overview of several common features of wired and wireless networking technologies comparing their frequency bands, data rates, and maximum number of nodes. 

Most popular communication technologies include CAN bus, Common Industrial Protocol (CIP), Ethernet, UPB, X10, Insteon, Z-wave, EnOcean, nanoNET, IEEE 802.15.4 (6LowPAN, Zigbee), IEEE~802.11 (Wi-Fi \cite{WLAN}), Bluetooth (Bluetooth Low Energy). The work in \cite{802.15.4vs802.11ah} investigates IEEE 802.15.4 against IEEE 802.11ah. The latter achieves better throughput than IEEE 802.15.4 in both idle and non-idle channels, although the energy consumption of IEEE 802.15.4 outperforms the values of IEEE 802.11ah, especially in dense networks. Furthermore, cellular networks include  WiMAX~\cite{estandarWiMAX, Wimax2} and 4G/5G LTE \cite{LTE}. In particular, 4G LTE uses carrier aggregation up to 100\,MHz, downlink~and uplink spatial multiplexing, multipath propagation and heterogeneous networks to provide higher throughput, greater bandwidth and efficiency, and facilitate simultaneous connection.
Highly integrated chipsets exist for most of these protocols, allowing for easy hardware integration. The~protocols mentioned have supporting development environments and in some cases manufacturers offer open source APIs.

The protocols presented offer at least some form of rudimentary congestion control, error recovery and some ad-hoc capability.
None of the communication protocols are designed for an actively hostile environment. 
Another specific technologies in use are \ac{RFID}, \ac{NFC} and~{UWB}.
 
\subsection{Enabling Protocols}
This subsection briefly addresses two main concerns: network routing and identification, and~RFID identification protocols. Regarding routing protocols, Routing Protocol for Low Power and Lossy Networks (RPL) is an IETF routing protocol based on IPv6 created to support minimal routing requirements through a robust topology (Point-to-Point (PtP), PMP).

On the other hand, the unique addresses follow two standards today, Ubiquitous ID and EPC Global. The EPC (Electronic Product Code) is a unique identification number stored on an RFID tag that is used basically in the supply chain management to identify items.
In order to decrease the number of collisions in the EPC Gen-2 protocol, and to improve tag identification procedure, researchers have proposed to use Code Division Multiple Access (CDMA) instead of the dynamic framed slotted ALOHA. A performance analysis of the RFID protocols in terms of the average number of queries and the total number of transmitted bits required to identify all tags in the system can be seen in \cite{PerformanceRFID}. 
The expected number of queries for tag identification  using the CDMA technique is lower that of the EPC Gen-2 protocol, because CDMA decreases the number of collisions.  However, when comparing the number of transmitted bits and the time to identify all tags in the system, EPC Gen-2 protocol performs better.
The EPC Global architectural framework is based on the EPC Information Service, which is provided by the manufacturer, and the ONS (Object Naming Service) that offers features similar to DNS (Domain Name Service). Being a central lookup service, the root of the ONS can be controlled or blocked by a company/country, unlike the DNS system. 

Identification methods, such as ubiquitous codes (uCode) and Electronic Product Codes (EPC), are not globally unique, although they provide a clear identity for each object within the network. Addressing methods of IoT objects, that include IPv4/IPv6,  assist to uniquely identify objects.

\subsection{Computation}
\vspace{-6pt}
 
\subsubsection{Hardware and Software Platforms} 
The growth of smart phone use in the last years has provided the basis for IoT hardware platforms. This tendency derives into new products being presented to the market at a fast pace. SoCs with very low power consumption, small form factor and oriented at supporting wireless communication technologies such as Wi-Fi and BLE, are being developed and enhanced. Arduino, Raspberry Pi, UDOO, FriendlyARM, Intel Galileo, Gadgetter, ESP8266, BeagleBone, Cubieboard, Zolertia Z1, WiSense, Mulle, and T-Mote Sky are just some examples of popular hardware platforms. Most of such devices are built on top of hardware solutions based on ARM Cortex M microcontrollers or ARM Cortex A microprocessors, but some use their own SoCs.

All these hardware platforms can be divided into two groups. On the one hand, there are SBCs (Single-Board Computers) like Raspberry Pi and Intel Galileo, which are powerful, and usually run some kind of modified Linux distribution. They support a vast set of security and communication alternatives, but their power consumption is high. On the other hand, the second type of platforms includes the motes. The ESP8266 or T-Mote Sky are good examples. They are much less power-hungry, being able to run on standard batteries for extended periods of time. However, they lack the processing capabilities of SBCs, and run on proprietary or ad-hoc software. In addition, one of the main problems of the currently available commercial motes is their lack of support for secure communication protocols and encryption. Nonetheless, motes recently presented address such an issue: for instance, the Arduino MKR1000 includes hardware acceleration for Elliptic Curve Cryptography (ECC), and the ESP32 has support for AES-256, SHA2, ECC and RSA-4096. 

Regarding software platforms, examples of Real-Time Operating Systems (RTOS) are Android, Contiki, TinyOS, LiteOS or Riot OS. The most common advanced programming environments and open standards are ARINC 653, Carrier Grade Linux, Eclipse, FACE, and POSIX. It must be also noted that Google and other important technological companies partnered with the auto industry to establish the Open Auto Alliance (OAA) \cite{OAA} to bring additional features to the Android platform to advance in the Internet of Vehicles paradigm.

\subsubsection{Cloud Platforms}
Connected devices need mechanisms to store, process, and retrieve data efficiently. However, the~amount of data collected in an IoT deployment may exceed the processing power of regular hardware and software tools. Moreover, IoT applications have to able to detect patterns or anomalies in the data when processing large amounts of data.

The emerging and developing technology of cloud computing is defined by the U.S. National Institute of Standards and Technology (NIST) as an access model to an on-demand network of shared configurable computing sources. Cloud computing enables researchers to use and maintain many resources remotely, reliably and at a low cost. The storage and computing resources of the cloud present the best choice for the IoT to store and process large amounts of data.
There are some platforms for big data analytics like Apache Hadoop and SciDB \cite{SurveyDataMining}.
The DoD is also trying to accelerate the adoption of commercial clouds \cite{DoDmemo}.
The cloud security model \cite{CloudSecRequirements} defines six information impact levels from 1 (public information) to 6 (classified information up to secret). As of May of 2015, there~were 26 Level~2 (Unclassified, Low-Impact) commercial cloud services approved with more on the way. Regarding Level 4/5 (Controlled Unclassified Information), there were one milCloud~\cite{milCloud} and one commercial cloud solution with more on the way. With respect to Level 6, there was one~milCloud.

In term of resources, besides the powerful servers in data centers, a lot of smart devices around us offer computing capabilities that can be used to perform parallel IoT data analytic tasks. Instead of providing applications specific analytics, IoT needs a common big data analytic platform which can be delivered as a service to IoT applications. Such an analytic service should not impose a considerable overhead on the overall IoT ecosystem.

\begin{figure}[H]
\centering
\includegraphics[width=0.8\columnwidth]{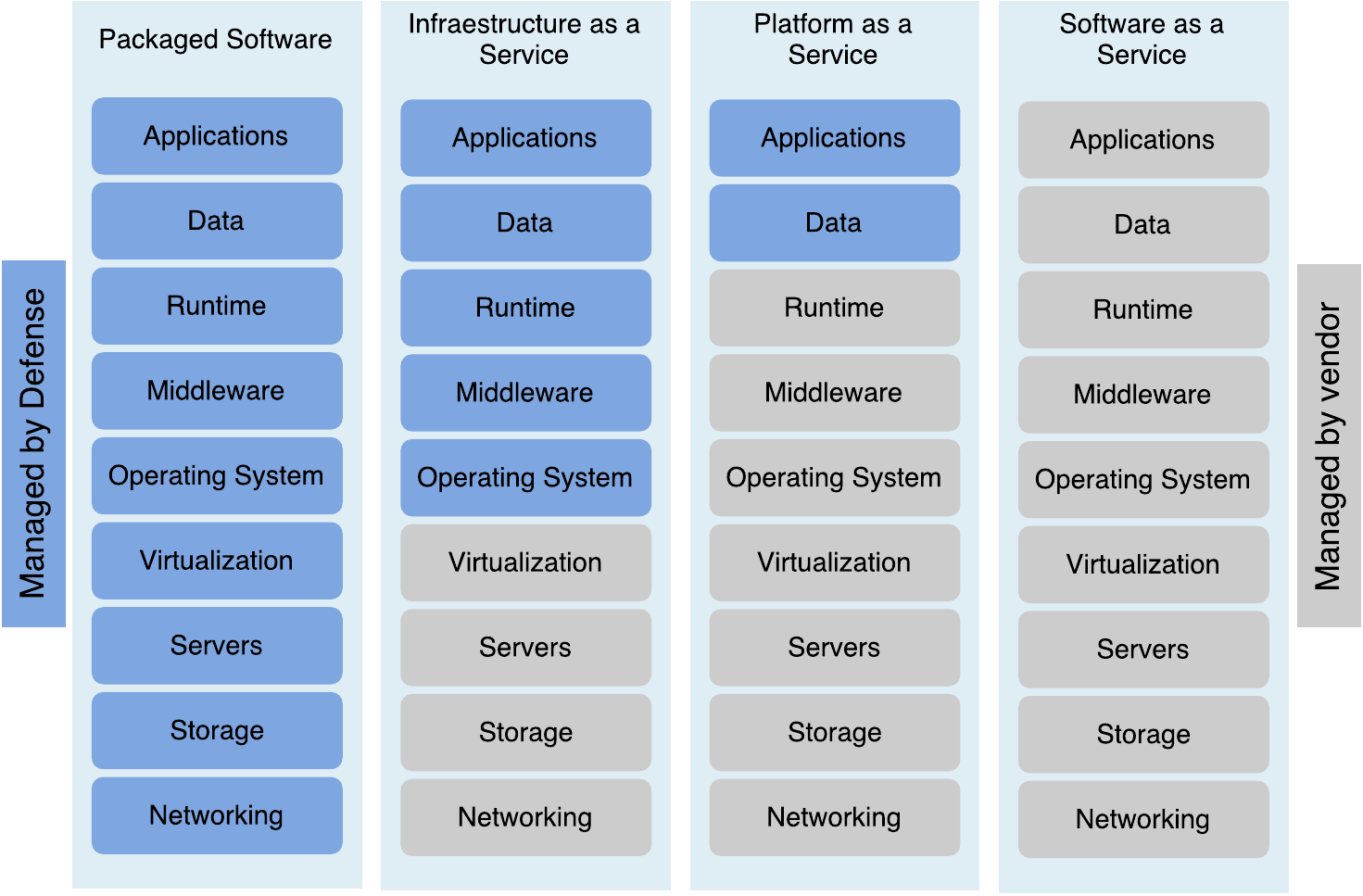}
\caption{Cloud paradigms: security inheritance and risks.}
\label{fig:SecurityInheritanceRisk}
\end{figure}

The three most popular cloud
paradigms are Infrastructure as a Service (IaaS), Platform as
a Service (PaaS), and Software as a Service (SaaS). Their structure and corresponding security risks are represented in Figure \ref{fig:SecurityInheritanceRisk}.

A scalable analytic service for time series data, Time Series analytics as a Service (TSaaaS) is presented in \cite{TSAaaS}. Pattern searching in TSaaaS can support effective searching on large amounts of time series data with very little overhead on the IoT system. 
TSaaaS is implemented as an extension to the Time Series Database service and it is accessible by RESTful web interfaces. Pattern searches are 10-100 times faster than other existing techniques, and the additional storage cost for the service provider accounts for only about 0.4\% of the original time series data.

Other feasible solution for IoT big data is to just keep track of the interesting data. Existing approaches include Principle Component Analysis (PCA), pattern reduction, feature selection, dimensionality reduction, and distributed computing methods \cite{SurveyDataMining}.

IoT can use numerous cloud platforms with different capabilities and strengths such as Google Cloud, AWS  \cite{AWS, CIAandAmazon}, Bluemix IoT Solutions \cite{Bluemix}, GENI, ThingWorx, OpenIoT, Arkessa, Axeda, Etherios, LittleBits...besides PS providers such as Avaya, Huawei Enterprise, West, or Microsoft. 

For example, Xively \cite{Xively} provides an open source PaaS solution for IoT application developers and service providers. It aims to securely connect devices to applications in real-time, it exposes accessible Application Programming Interfaces (APIs), and it provides interoperability with many protocols and environments. It enables the integration of devices with the platform by libraries and facilitates communication via HTTP(s), Websocket, or MQTT. It integrates with other platforms using Python, Java, and Ruby libraries; and distributes data in numerous formats such as JSON, XML and CSV \cite{Verma}. It also allows users to visualize their data graphically and to remotely control sensors by modifying scripts to receive and send alerts. It is supported by many Original Equipment Manufacturers (OEM) like Arexx, Nanode, OpenGear, Arduino or mBed.

Nimbits \cite{Nimbits} connects smart embedded devices to the cloud, it performs data analytics and generates alerts. Moreover, it connects to websites and can store, share and retrieve sensor's data in various formats, including text based, numeric, GPS, JSON or XML. It uses XMPP to exchange data or messages. The core is a server that provides REST web services for logging and retrieving raw and processed data.

The authors of \cite{Mazhelis} summarize some of the characteristics of a number of available cloud platforms. The metrics include: support of  gateway devices to bridge the short range network and wide area network, support of discovery, delivery, configuration and activation of services, provision of a proactive and reactive assurance of platform, support of accounting and billing of services, and, finally, support of standard application protocols.
All the platforms analyzed by the authors support sensing or actuation devices, a user interface to interact with them, and a web component to run the business logic of the application on the cloud. None of such platforms supports the DDS protocol. 

Voegler et al. \cite{Voegler} propose a novel infrastructure to provide application packages on resource-constrained heterogeneous edge devices elastically in large-scale IoT deployments. It~enables push-based (commands down to the tactical units) as well as pull-based (from ground to decision-making) deployments supporting different topologies and infrastructure requirements.

The efficient use of cloud based resources requires the previous selection of software architectures for both communications and processing. Centralized cloud approaches, in which raw data is transmitted to the cloud for analysis, are non-viable in tactical environments, as some of the main restrictions in military IoT scenarios are time and resources. 
Furthermore, even if a device has a high-bandwidth link to a local resource, it is not likely that all devices will have good connectivity to the same cloud-based platform. Thus, relying on tactical wireless networks, any approach that requires a centralized cloud infrastructure is not likely to work properly. Moreover, in a centralized cloud infrastructure, processing represents a complex and computationally expensive procedure, which~leverages sophisticated big data tools. Finally, there is a significant delay between the time of the IoT data generation and when the results become available.

In order to address the issue of distributed infrastructures for IoT data analysis, researchers have started investigating distributed cloud architectures. The idea consists in extending and complementing a small number of large cloud data centers located in the core of the network, where~most computational and storage resources are concentrated, with a large number of tiny cloud data centers located at the boundary between the wired Internet and the IoT. This would enable data analysis applications to benefit from the elastic nature of cloud-based resources while pushing the computation closer to the IoT, with obvious advantages in terms of reducing communications overhead and processing times. There is also research to support the processing of raw IoT data close to the source of their generation, particularly, the processing and filtering of raw IoT data and the exploitation of IoT specific computational solutions for data analysis purposes. Several proposals have emerged from the realization that not all the raw data generated are equally important, and that applications might be better served by focusing only on important data. 
The \ac{QoI} and \ac{VoI} concepts arise to extend Shannon's information theory to consider both the probabilistic nature of the uncertainties. These efforts are highly relevant for the military IoT, as~the processing and exploitation of the information is made according to the utility for its users. Thus, the~ability of supporting the user in more effective decision making has potential to reduce the amount of computational and bandwidth resources required for data analysis and dissemination.

Emerging hardware and computational solutions for embedded platforms require new software architectures to fulfill their potential. For instance, neuromorphic processors, hybrid CPUs/FPGAs processors feature programming models that are different from the ones of the server CPUs typical of cloud data centers.

\subsubsection{Fog Computing}

\begin{figure}[H]
\centering
\includegraphics[width=0.5\columnwidth]{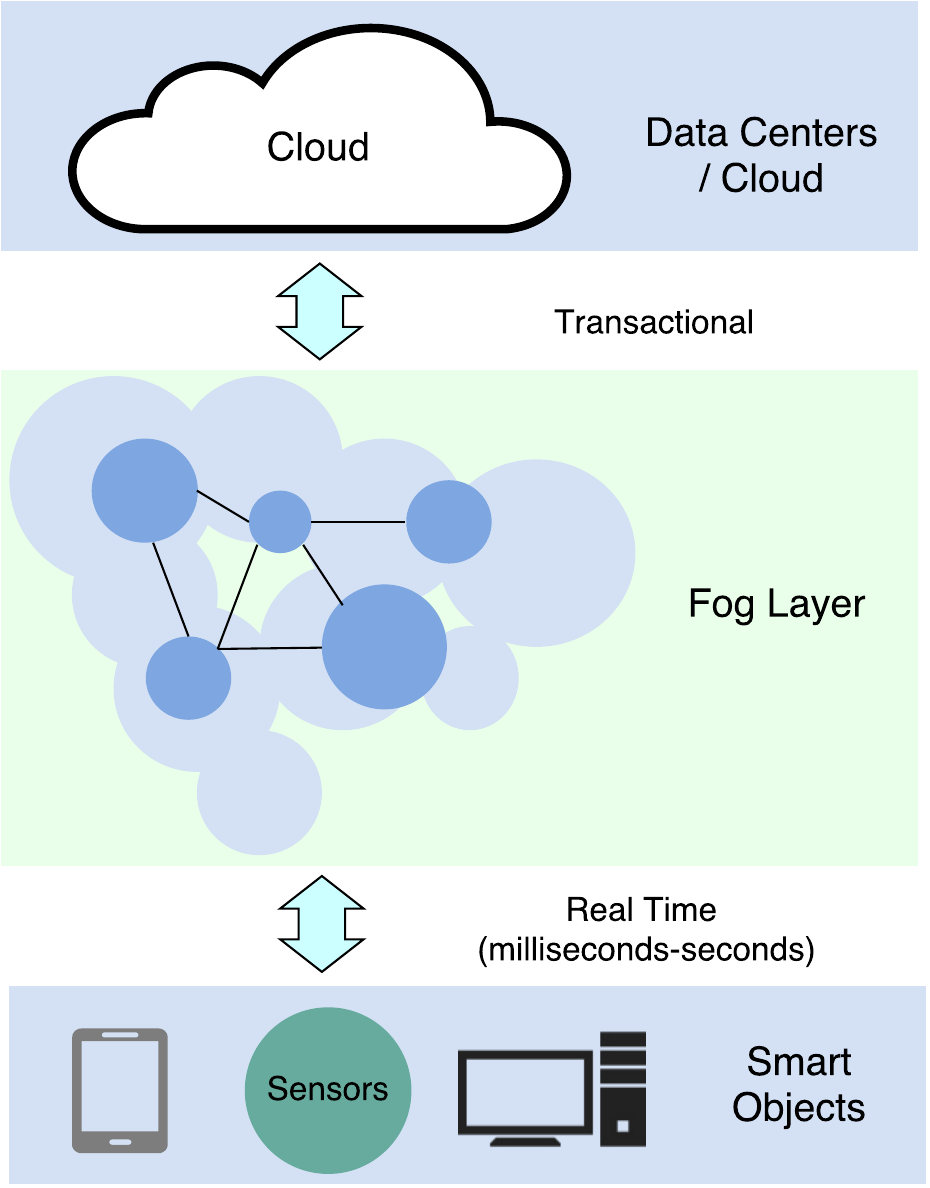}
\caption{Fog computing paradigm.}
\label{fig:FogComputingParadigm}
\end{figure}

Several research concepts, such as fog computing, cloudlets, mobile edge computing and IoT-centric clouds have been recently proposed to complement the distributed cloud architectures for IoT data analysis and security \cite{virtualization, ciphercloud, cloudlets, CloudSurveillance}.  Fog computing has the potential to increase the overall performance of IoT applications as it tries to perform part of high level services, which are offered by the cloud, inside local resources. This paradigm is depicted in Figure \ref{fig:FogComputingParadigm}.

So far researchers have focused mostly on how to extend elastic resource consumption paradigms and big data solutions to distributed cloud configurations, instead of proposing new methodologies, paradigms and tools to efficiently exploit the capabilities of IoT hardware. Fog computing can act as a bridge between smart devices and large-scale cloud computing and storage services. Because of their proximity to the end-users, it has the potential to offer services faster. 
There is a significant difference in scale between the fog and the cloud: the latter has massive computational, storage, and communications capabilities compared to the former. Mobile network operators are potential providers of fog computing since they can offer fog services like IaaS, PaaS, or SaaS at their service network or even at a cell tower, or even a type of transversal service, that is IoT as a Service (IoTaaS).
The main advantages of fog computing can be summarized as follows:
\begin{itemize}[leftmargin=*,labelsep=5mm]
\item Location: fog resources provide less delay because they are positioned between smart objects and the cloud data-centers.
\item Distribution: it is possible to deploy many of such micro centers close to the end-users as their cost is a small fraction of a cloud data-center.
\item Scalability: the number of micro fog centers can be increased to cope with the increasing load and the increased number of end-users.
\item Density of devices: fog helps to provide resilient and replicated services.
\item Mobility support: fog resources act as a mobile cloud.
\item Real-time: it provides better performance for real-time interactive services.
\item Standardization: fog resources can interoperate with various cloud providers. 
\item On-the-fly analysis: fog resources can perform data aggregation to send partially processed data as opposed to raw data to the cloud data center for further processing.
\end{itemize}

One important aspect of cloud platforms is the ability to interact with different application protocols. 
IoTCloud \cite{Fox} is a project aimed at providing a scalable and high performance cloud platform for IoT applications. 

Fog computing still needs research to resolve other issues like reliability, mobility and security of analytical data on the edge devices. Chang et al. \cite{Chang} presented a fog computing model that brings information-centric cloud capabilities to the edge to deliver services with reduced latency and bandwidth.
This situation calls for the need of a better horizontal integration between different application layer protocols. Several attempts of integration have been made in recent literature. For~example,  Ponte \cite{Ponte} offers uniform open APIs to enable the automatic conversion between various IoT applications protocols such as HTTP, CoAP, and MQTT.
Nevertheless, the capability to perform any-to-any automatic protocol conversion implies that the underlying packet communication tends to be more verbose in order to be application agnostic. Furthermore, Ponte, as many other protocol gateways, assumes the underlying devices to be TCP/IP enabled. Also, resource-constrained devices are not considered at all in this solution.

The Kura project \cite{Kura} is an M2M application platform that provides a Java/OSGi-based container for applications running in service gateways. Kura covers I/O access, data services, watchdog, network configuration and remote management. Scada \cite{Scada} focuses on providing a way to connect different industrial devices to a common communication system. It aims to ease post-process and data visualization. Another Eclipse IoT project is Krikkit \cite{Krikkit} that comes with a RESTful API that allows a developer to acquire data in edge devices such as sensor gateways. It uses a publish/subscribe model by which data acquiring rules or policies are registered on edge devices. Lin et al. \cite{Lin} use Software Defined Radio (SDR) technology as part of their platform to sense and transform the wireless signals in the frequency spectrum.
An approach based on software-defined networking (SDN) for IoT tasks is presented in \cite{Qin}. They develop a middleware with a layered SDN controller to manage dynamic and heterogeneous multi-network environments.

Therefore, a new intelligent IoT gateway is needed. An intelligent ruled-based gateway should allow programmers to control the wire protocol traffic to optimize performance based on the specific needs of the given application \cite{Al-Fuqaha}.
The rules that pertain to autonomic management and data aggregation services can result in the generation and transmission of new data and management packets by the gateway itself. It will result in a lighter protocol stack that relies on uIP/lwIP (with no need for TCP/UDP, TLS or other security protocols on resource-constrained devices).

On resource-rich devices that support the TCP/IP protocol suite, IoT applications are implemented on top of a variety of application level protocols and frameworks including REST, CoAP, MQTT, and~AMQP, among others. On the other hand, devices that do not have the  resources required to support TCP/IP have a myriad of interoperability issues that limit the potential applications of the IoT.

{ The fulfillment of complex requirements such as ubiquity, scalability and high-performance lead to a convergence between the IoT and cloud through federation and multi-cloud architectures. Cloud federation is one of the core concepts for the design, the deployment and the management of decentralized edge cloud infrastructures. Since federated systems inherit all of the fundamental aspects of distributed computing, they can certainly leverage many existing standards that have been developed in this arena over the past years.

The near-future evolution of IoT clouds is discussed in \cite{Celesti}. There the authors describe a three-stage evolution towards the creation of an IoT federation. The first of such stages is called monolithic and involves embedded devices that would be connected to IoT cloud systems to provide basic IoTaaS (the services would be developed either with stand-alone pieces of software or by means of container virtualization technology). The next stage is named vertical supply chain. In such a stage the IoT cloud providers leverage IoTaaS offered by other providers. Finally, the third stage corresponds to the real IoT cloud federation, where IoT cloud providers will federate to extend their sensing capabilities, adopting the container virtualization technology massively in order to create more flexible IoTaaS.

Likewise, numerous research projects and initiatives \cite{ClouT, MUSA} are focused on the realization of innovative architectures for the Cloud-IoT, enabling features such as autonomous service provisioning and management. Indicatively, such a concept may be applicable to 5G technological solutions \cite{Chen2015}, such as SDN. For example, cloud-based mechanisms will enable the incorporation of resources and services independent of their location across distributed computing and data storage infrastructures. The challenge will be the integration of these different standardized capabilities into a coherent end-to-end federation model. According to the cloud federation's organization, access and scale, six federation deployment models can be identified \cite{CloudFederation}: simple pairwise federation, hierarchical federation, peer-to-peer federations, brokers or interclouds.}

The main challenges of employing cloud computing for the mission-critical IoT include the synchronization to provide real-time services (since they are built on top of various cloud platforms); the need for a balance between cloud service environments and IoT requirements, considering the differences in infrastructure; and to solve issues like the lack of standardization, the complicated management and the enhancement of the reliability and the security. Hashizume et al. \cite{Hashizume2013SecurityCloud} provide an analysis of vulnerabilities, threats and countermeasures in the cloud considering the three service delivery models: SaaS, PaaS and IaaS. The article ends emphasizing the need for new security techniques (such as firewalls, Intrusion Detection System (IDS), Intrusion Prevention System (IPS) and data protection) as well as the redesign of traditional cloud solutions.

There are two main security challenges in the cloud-centric IoT: secure storage and authorized data sharing in near real time. Authentication prevents access by illegitimate users or devices, and it can prevent legitimate devices from accessing resources in an unauthorized way. Scalable authentication schemes have been widely studied for traditional computer networks as well as WSNs. Cloud-centric authentication as a service has also been considered to minimize task overhead on user devices. For~example, Butun et al. \cite{PScloud} present a hierarchical authentication as a service for PS networks. The~proposed lightweight cloud-centric multi-level framework addresses scalability for IoT-worn devices.
In the proposed CMULA scheme, public safety responders and devices are authenticated through the Cloud Service Provider (CSP). This approach enables
easier mobility management.
The~network consists of four entities: users (the chief officers who are registered in the emergency system and are responsible for managing the responders on site), wearable nodes, a Wearable Network Coordinator (WNC) (responsible for managing all sensors attached to the responder’s body), and a CSP that serves as certification authority for the IoT-based public safety network. It considers a public key infrastructure (PKI) issuing ECC throughout the cloud-centric IoT. Elliptic Curve Digital Signature Algorithm (ECDSA), a variant of ECC, is used for digital certificate generation and verification. Another variant of ECC, the Elliptic Curve Diffie-Hellman (ECDH) key exchange algorithm  is used to exchange the secret message authentication code (MAC) keys in the initialization phase. Once~a user is authenticated to a CSP, wearable devices can be accessed through a WNC.
Other existing studies on cloud security have been focused on issues concerning cloud security, identity management, and~access control or architecture layers. For example, Li et al. \cite{MobileCloud} review mechanisms and open issues for mobility-augmented service provisioning. As a result, they discover open challenges with respect to overhead, heterogeneity, QoS, privacy and security.
Authors in \cite{CloudSecurity} provide an integrated solution to cloud security based on the so-called Cloud Computing Adoption Framework (CCAF) framework. It protects data security and predicts the probable consequences of abnormal situations by using Business Process Modeling Notation (BPMN) simulations. The multi-layer description of CCAF is as follows. The first layer is for access control: a firewall allows the access just to certain members. The~second layer consists of the IDS/IPS to provide up-to-date technologies to prevent attacks such as DoS, anti-spoofing, port scanning, pattern-based attacks, parameter tampering, cross~site scripting, SQL injection or cookie poisoning. The identity management is enforced to ensure that the right level of access is only granted to the right person. Finally, the third layer is convergent encryption. The~results of CCAF expose real-time protection of all the data, blocking and quarantining the majority of the threats.

\subsection{Digital Analytics}
Analytical software manage the excessive volume of data that needs to be transferred, stored and analyzed. It would require flexible acquisition processes by the governments to integrate cutting-edge technologies quickly.
Many applications would depend on real-time analysis to enable automated responses.
Other system would process data into simple interfaces that allow humans to leverage big data in convenient ways.

Semantic Web technologies have been acknowledged as important to support for data integration, reasoning and content discovery \cite{SemanticWebC2}. Particularly, three established elements have been identified as desirable IoT tactical capabilities:
\begin{itemize}[leftmargin=*,labelsep=5mm]
\item Open Integration standards: they facilitate interoperability among devices with different capabilities and ownership through supporting ontologies. IoT ontologies should be integrated with existing community standards.
\item Reasoning support: Ontology-based reasoning has been applied towards military sensor management systems, including those tasked with pairing sensors to mission tasks.
Gomez~et~al.~\cite{Gomez2008} present an ontology based on Military Missions and Means Framework that formalizes sensor specifications as well as expressing corresponding task specifications. 
When~there is limited network connectivity, such reasoning capabilities could be applied tocontinually assess how available IoT resources can be utilized.
\item Data Provenance: the steps taken to generate data have been commonly acknowledged as important towards assessment of data quality and trustworthiness. 
In a military context, issues of provenance will be a dominant concern because the state, ownership, and reliability of devices will be uncertain. The capability will be critical when automated or semi-automated content assessment becomes desirable. New architectures will need to incorporate provenance and trust management tightly integrated in IoT technologies. The W3 PROV specification \cite{W3PROV} is a primary standard for digital provenance representation, which is now being extended for IoT.
\end{itemize}

\section{Main Challenges and Technical Limitations} \label{Challenges}

There are significant challenges in the development and deployment of existing and planned military IoT systems. Nowadays, only a small number of military systems leverage the full advantages of IoT. Ongoing NATO Research Task Group (RTG) 'Military Applications of Internet of Things'  (IST-147) is examining a number of  critical issues identified by the recommendations from two previous exploratory team activities: IST-ET-076, 'Internet of Military Things' which examined topics relevant to the application of IoT technologies, and IST-ET-075, 'Integration of Sensors and Communication Networks', which addressed networking issues.
The deployment of IoT-related technologies is in segregated vertical stovepipes making it difficult to secure them, and limiting the ability to communicate across systems and generate synergies from different data sources. Main~defense concerns include the dependence of manual entry, the limited processing of data, the lack of automation, and the fragmented IT architecture \cite{DoDplan}.
 
Furthermore, nowadays the military does not have sufficient network connectivity on the battlefield to support broader IoT deployment. It will require key investments in several technical enablers  according to its information value loop \cite{Deloitte}. The roadmap for near-future research and technology developments is  depicted in Table \ref{table:Roadmap}.

\vspace{-18pt}
 \begin{footnotesize}  
 \begin{center}
 \begin{longtable}{p{2.5cm}p{12.8cm}c} 
 
\caption{{Roadmap for technologies and ongoing research.}}\label{table:Roadmap}\\ \toprule

\multicolumn{1}{c}{\textbf{Research}} & \multicolumn{1}{c}{\textbf{Timeframe 2016--2020}}   \\
 \midrule
%\endfirsthead
%\hline
%\multicolumn{2}{c}%
%{{\bfseries \tablename\ \thetable{} -- continued from previous page}} \\
%\hline \multicolumn{1}{c}{\textbf{Research}} & \multicolumn{1}{c}{\textbf{Timeframe 2016-2020}}  \\ \hline 
%\endhead
%
%\hline \multicolumn{3}{r}{{Continued on next page}} \\ \hline
%\endfoot
%
%\hline \hline
%\endlastfoot
\enlargethispage{0.5cm}
\multirow{9}{*}{Identification}
& $\bullet$ Identity management\\
& $\bullet$ Open framework for the IoT\\
& $\bullet$ Soft Identities\\ 
& $\bullet$ Semantics\\ 
& $\bullet$ DNA identifiers\\
& $\bullet$ Convergence of IP and IDs and addressing scheme: unique or multiple IDs\\
& $\bullet$ Extend the ID concept (more than ID number)\\
& $\bullet$ Electro Magnetic Identification (EMID)\\
& $\bullet$ Multi methods, one ID\\
 \midrule
\multirow{8}{*}{Architecture}
& $\bullet$ Network of networks architectures\\ 
& $\bullet$ Adaptive and context based architectures\\ 
& $\bullet$ Self-managing properties (they include self-configuring, self-healing, self-optimizing, self-protecting, self-awareness, self-adaptation, self-evolving and self-anticipating)\\
& $\bullet$ Cognitive and experimental architectures\\
& $\bullet$ Code in tags to be executed in the tag or in trusted readers with global applications, adaptive coverage, universal authentication of objects, recovery of tags following power loss, more memory, less energy consumption, 3-D real time location/position embedded systems\\
& $\bullet$ Cooperative position cyber-physical systems\\
 \midrule
\multirow{3}{*}{Infrastructure}
& $\bullet$ Cross domain application deployment\\ 
& $\bullet$ Integrated IoT, multi-application and multi-provider infrastructures\\ 
& $\bullet$ General purpose IoT: global discovery mechanism\\  
 \midrule
\multirow{7}{*}{Applications}
& $\bullet$ IoT device with strong processing and analytics capabilities\\ 
& $\bullet$ Handling heterogeneous high capability data collection and processing\\  
& $\bullet$ Application domain-independent abstractions and functionality\\
& $\bullet$ Cross-domain integration and management\\
& $\bullet$ Context-aware adaptation of operation\\
& $\bullet$ Standardization of APIs\\
& $\bullet$ Mobile applications with bio-IoT-human interaction\\
 \bottomrule
 \end{longtable}
\end{center}

\newpage

\setcounter{table}{1}

  \begin{center}\vspace{-48pt}
  \footnotesize

 \begin{longtable}{p{2.5cm}p{12.8cm}c}
% \caption{{\em Cont.}} \label{xx}
%\multicolumn{1}{c}{\textbf{Research}} & \multicolumn{1}{c}{\textbf{Timeframe 2016--2020}}   \\
 %\midrule
  
\caption{{\em Cont.}}\label{table:2}\\ \toprule

\multicolumn{1}{c}{\textbf{Research}} & \multicolumn{1}{c}{\textbf{Timeframe 2016--2020}}   \\
 \midrule
\multirow{11}{*}{Communications}
& $\bullet$ Wide spectrum and spectrum aware protocols\\ 
& $\bullet$ Ultra-low power system on chip, multi-protocol chips\\
& $\bullet$ Multi-functional reconfigurable chips\\
& $\bullet$ On-chip antennas\\
& $\bullet$ On-chip networks and multi-standard RF architectures\\
& $\bullet$ Seamless networks\\
& $\bullet$ Gateway convergence\\
& $\bullet$ Hybrid network technologies convergence\\
& $\bullet$ 5G developments\\
& $\bullet$ Collision-resistant algorithms\\
& $\bullet$ Plug-and-play tags, self-repairing tags\\
 \midrule
\multirow{9}{*}{Network}
& $\bullet$ Self-aware, self-configuring, self-learning, self-repairing and self- organizing~networks\\ 
& $\bullet$ Sensor network locations transparency\\
& $\bullet$ IPv6-enabled scalability\\
& $\bullet$ Ubiquitous IPv6-based IoT deployment\\
& $\bullet$ Software defined networks\\
& $\bullet$ Service based network\\
& $\bullet$ Multi authentication, integrated/universal authentication\\
& $\bullet$ IPv6-based Internet of Everything (smart cities)\\
& $\bullet$ Robust security based on a combination of ID metrics\\
 \midrule
%\pagebreak
\multirow{14}{*}{Software}
& $\bullet$ Goal oriented: distributed intelligence, problem solving, Things-to-Things collaboration environments\\
& $\bullet$ IoT complex data analysis\\
& $\bullet$ IoT intelligent data visualization\\ 
& $\bullet$ Hybrid IoT\\
& $\bullet$ User oriented: the invisible IoT, things-to-Humans collaboration, IoT 4 All and User-centric IoT\\
& $\bullet$ Quality of Information and IoT service reliability\\
& $\bullet$ Highly distributed IoT processes\\
& $\bullet$ Semi-automatic process analysis and distribution\\
& $\bullet$ Fully autonomous IoT devices\\
& $\bullet$ Micro operating systems\\
& $\bullet$ Context aware business event generation\\
& $\bullet$ Interoperable ontologies of business events\\
 \midrule
\multirow{4}{*}{Signal Processing}
& $\bullet$ Context aware data processing and data responses\\ 
& $\bullet$ Distributed energy efficient data processing\\  
& $\bullet$ Cognitive processing and optimization\\  
& $\bullet$ Common sensor ontologies (cross domain)\\ 
 \midrule
\multirow{5}{*}{Discovery}
& $\bullet$ Automatic route tagging and identification management centers\\
& $\bullet$ Semantic discovery of sensors\\ 
& $\bullet$ Cognitive search engines\\
& $\bullet$ Autonomous search engines\\
& $\bullet$ Scalable Discovery services for connecting things with services while respecting security,
privacy and~confidentiality\\
 \midrule
\multirow{7}{*}{Energy efficiency}
& $\bullet$ Energy harvesting (biological, chemical, induction)\\
& $\bullet$ Power generation in harsh environments\\
& $\bullet$ Biodegradable batteries\\
& $\bullet$ Nano-power processing unit\\
& $\bullet$ Energy recycling\\
& $\bullet$ Long range wireless power\\
& $\bullet$ Wireless power everywhere, anytime\\
\bottomrule
\end{longtable}\vspace{36pt}
\end{center}
\setcounter{table}{1}
  \begin{center}
  
 \begin{longtable}{p{2.5cm}p{12.8cm}c}
% \caption{{\em Cont.}} \label{xx}
%\multicolumn{1}{c}{\textbf{Research}} & \multicolumn{1}{c}{\textbf{Timeframe 2016--2020}}   \\
 %\midrule
  
\caption{{\em Cont.}}\label{table:3}\\ \toprule

\multicolumn{1}{c}{\textbf{Research}} & \multicolumn{1}{c}{\textbf{Timeframe 2016--2020}}   \\
 \midrule
\multirow{16}{*}{Security}
& $\bullet$ Low cost, secure and high performance identification/authentication devices\\
& $\bullet$ User centric context-aware privacy\\
& $\bullet$ Privacy aware data processing\\
& $\bullet$ Security and privacy profiles and policies\\
& $\bullet$ Context centric security\\
& $\bullet$ Homomorphic Encryption, searchable Encryption\\
& $\bullet$ Protection mechanisms for IoT DoS/DdoS attacks\\
& $\bullet$ Self-adaptive security mechanisms and protocols\\
& $\bullet$ Access control and accounting schemes\\
& $\bullet$ General attack detection and recovery/resilience\\
& $\bullet$ Cyber Security\\
& $\bullet$ Decentralized self-configuring methods for trust establishment\\
& $\bullet$ Novel methods to assess trust in people, devices and data\\
& $\bullet$ Location privacy preservation\\
& $\bullet$ Personal information protection from inference and observation\\
& $\bullet$ Trust Negotiation\\
 \midrule
\multirow{4}{*}{Interoperability}
& $\bullet$ Automated self-adaptable and agile interoperability\\ 
& $\bullet$ Reduced cost of interoperability\\  
& $\bullet$ Open platform for IoT validation\\
& $\bullet$ Dynamic and adaptable interoperability for technical and semantic areas\\
\hline
\multirow{6}{*}{Standardization}
& $\bullet$ M2M standardization \\ 
& $\bullet$ Standards for cross interoperability with heterogeneous networks\\ 
& $\bullet$ Standards for IoT data and information sharing\\
& $\bullet$ Standards for autonomic communication protocols\\
& $\bullet$ Interaction standards\\
& $\bullet$ Behavioral standards\\
 \midrule
\multirow{16}{*}{Hardware}
& $\bullet$ Smart bio-chemical sensors\\ 
& $\bullet$ Nano-technology and new materials\\
& $\bullet$ Interacting/Collaborative tags\\
& $\bullet$ Self-powering sensors\\
& $\bullet$ Polymer based memory, ultra-low power EPROM/FRAM\\
& $\bullet$ Molecular sensors\\
& $\bullet$ Transparent displays\\
& $\bullet$ Biodegradable antennas\\
& $\bullet$ Nano-power processing units\\
& $\bullet$ Biodegradable antennas\\
& $\bullet$ Multi-protocol frontends\\
& $\bullet$ Collision free air to air protocol and minimum energy protocols\\
& $\bullet$ Multi-band, multi-mode wireless sensor architectures implementations\\
& $\bullet$ Reconfigurable wireless systems\\
& $\bullet$ Micro readers with multi-standard protocols for reading sensor and actuator data\\
& $\bullet$ System-in Package (SiP) technology including 3D integration of components\\
 \bottomrule
\end{longtable}
\end{center}
\end{footnotesize}

 \vspace{-18pt}
%\pagebreak

As it can be seen in  Table \ref{table:Roadmap}, security is the most significant demand for IoT adoption across the military. Defense faces a large number of simple devices and applications with unique vulnerabilities for electronic and cyber warfare. 
Data analytics and process capacity are additional limiting factors.

\subsection*{From COTS to Mission-Critical IoT: Further Recommendations}

Despite the ongoing technological research, the following recommendations were obtained from the analysis of the previous sections:

\begin{itemize}[leftmargin=*,labelsep=5mm]

\item Introduce rapid field testing: the military should consider creating a dedicated technology comprising military personnel in a live training environment to experiment with technologies and get real end-user feedback early in the development process. This testbed could change the way the military accomplishes its mission, or introduces creative new ways to use IoT devices and applications. Its goal would be twofold: to recognize devices and systems with potential applications and, second, to identify completely new strategies, tactics, and methods for accomplishing missions using COTS. 

\item The military can to a certain extent, take advantage of civilian mobile waveforms such as 4G/5G LTE \cite{ICMCIS}. Nevertheless, those advances will need to be paired with military-specific communications architectures (e.g., multiband radios with scarce bandwidth, MANET topologies and defensive countermeasures).
\item Use \ac{PaaS} to deliver web-based services without building and maintaining the infrastructure, thereby creating a more flexible and scalable framework to adjust and update the systems. Adopting PaaS also carries risks for the military, and requires private contractors to implement additional security procedures.
\item Realize a comprehensive trust framework that can support all the requirements of IoT for the military. Many state-of-the-art approaches that address issues such as trust and value depend on inter-domain policies and control. In military environments, policies will likely be contextual and transient, conflated by inter-organizational and adversarial interactions.
\item Information theories will need to focus on decision making and cognitive layers of information management and assimilation. Further, methods for eliciting causal relationships from sparse and extensive heterogeneously-sourced data will require additional theoretical research.
\end{itemize}

There are key enabling technologies in which governments and defense can invest today to enable greater IoT deployment in the near future.
Besides, the adoption of IoT will require the compromise of all stakeholders. 
Another constraint is the current budget environment \cite{SpecialOperationsBudget2015, F-35}: defense is reluctant to spend limited budgets on up-front costs for generating significant future savings. Defense should adopt new ways to access innovation, adopting commercial best practices for technology development and acquisition. An enhanced collaboration with the private sector is needed to field and update IoT systems with cutting-edge technology. Cultural differences between defense and innovators of the private sector as well as intellectual property and export restrictions, discourage companies from collaborating with the military. Also companies and innovators may see little benefit in catering the complex and demanding operational requirements of defense and public safety, which~is a small and demanding customer in comparison to commercial markets.  Creating affordable and high-value systems that deliver enhanced situational awareness for military has a proven business value. Complementing this intelligence with integrated commercial IoT data is also a compelling business model for innovative defense and public safety contractors and system integrators.

 %%%%%%%%%%%%%%%%%%%%%%%%%%%%%%%%%%%%%%%%%%

\section{Conclusions} \label{SecCon}

This article examined how the defense industry can leverage the opportunities created by the commercial IoT transformation. Main topics relevant to the application of IoT concepts to the military and public safety domain were explained. In order to perform the study, we propose significant scenarios such as: C4ISR, fire-control systems, logistics (fleet management and individual supplies), smart cities operations, personal sensing, soldier healthcare and workforce training, collaborative and crowd sensing, energy management, and surveillance. The added value and the risk of applying IoT technologies in the selected scenarios were also assessed. Based on the operational requirements, we~proposed the architecture, technologies and protocols that address the most significant capabilities.

Commercial IoT still faces many challenges, such as standardization, scalability, interoperability, and security. 
Researchers working on defense have to cope with additional issues posed by tactical environments, and the nature of operations and networks. There are three main differences between Defense/PS IoT and COTS IoT: the complexity of the deployments, the resource constraints (basically the ones related to power consumption and communications), and the use of centralized cloud-based~architectures.

Organic transitions such as supply chain management and logistics will naturally migrate to mission-critical environments.
Beyond the  earliest military IoT innovations, complex battlefields will require additional research advances to address the specific demands. In addition to addressing various technical challenges, this work identified vital areas of further research in the 2016--2020 timeframe. 
Moreover, battlefield domains that closely integrate human cognitive processes will require new or extensions of current theories of information that scale into deterministic situations.

We can conclude that broader deployment of defense and PS IoT applications will take time. Nevertheless, there are areas where governments and defense can generate significant savings and advantages using existing \ac{COTS} technologies and business practices. Defense and PS needs to adopt best practices for technology development and acquisitions from the private sector, and~should consider a bottom-up model of innovation and procurement. As in any industry, there is no one-size-fits-all solution to the IoT for defense. The military and first responders should establish a testbed for identifying and experimenting with technologies that could remodel the way missions are accomplished, and which would serve as a link between warfighters in the field and IoT developers. The military should invest in developing new security techniques that can be applied to COTS devices and applications, including those hosted in the cloud.
The focus should be on investing in scalable security measures instead of securing individual systems.
This approach will give defense and PS greater leverage in their IoT investments, allowing them better returns per dollar spent on proprietary R\&D while exploiting the military IoT potential.

\vspace{6pt}
%%%%%%%%%%%%%%%%%%%%%%%%%%%%%%%%%%%%%%%%%%

\acknowledgments{
This work has been funded by the Spanish Ministry of Economy and Competitiveness under grants TEC2013-47141-C4-1-R and TEC2015-69648-REDC.}

%%%%%%%%%%%%%%%%%%%%%%%%%%%%%%%%%%%%%%%%%%

%\authorcontributions{\textbf{Author Contributions:} For research articles with several authors, a short paragraph specifying their individual contributions must be provided. The following statements should be used ``X.X. and Y.Y. conceived and designed the experiments; X.X. performed the experiments; X.X. and Y.Y. analyzed the data; W.W. contributed reagents/materials/analysis tools; Y.Y. wrote the paper.'' Authorship must be limited to those who have contributed substantially to the work reported.}

\authorcontributions{
Paula Fraga-Lamas, Tiago M. Fern\'andez-Caram\'es and Manuel Su\'arez-Albela
contributed to the overall study design, data collection and analysis,
and writing of the manuscript. Luis Castedo and Miguel~Gonz\'alez-L\'opez contributed to
the overall writing of the manuscript. All of the authors approved the final version of the manuscript.}
%%%%%%%%%%%%%%%%%%%%%%%%%%%%%%%%%%%%%%%%%%

%\conflictofinterests{\textbf{Conflicts of Interest:} Declare conflicts of interest or state ``The authors declare no conflict of interest.'' Authors must identify and declare any personal circumstances or interest that may be perceived as inappropriately influencing the representation or interpretation of reported research results. Any role of the funding sponsors in the design of the study; in the collection, analyses or interpretation of data; in the writing of the manuscript, or in the decision to publish the results must be declared in this section. If there is no role, please state ``The founding sponsors had no role in the design of the study; in the collection, analyses, or interpretation of data; in the writing of the manuscript, and in the decision to publish the results''.} 

\conflictofinterests{The authors declare no conflict of interest. The founding sponsors had no role in the design of the study; in the collection, analyses, or interpretation of data; in the writing of the manuscript, and in the decision to publish the results.} 

%%%%%%%%%%%%%%%%%%%%%%%%%%%%%%%%%%%%%%%%%%
%% optional

\abbreviations{The following abbreviations are used in this manuscript:\\

\noindent\hspace{-0.65em}\begin{tabular}{ll}
 DoD & Department of Defense \\
ACV & Armored Combat Vehicles \\
AJ & Anti-Jamming \\
BFT & Blue Force Tracking \\
C2 & Command and Control \\
C4ISR & Command, Control, Communications, Computers, Intelligence, Surveillance and Reconnaissance \\
COP & Common Operational Picture\\
COTS & Commercial Off-The-Shelf\\
EPC & Electronic Product Code\\
EPM & Electronic Protection Measures\\
HAP & High-Altitude Platforms\\
IaaS & Infrastructure as a Service\\
IoT & Internet of Things\\
ISR & Intelligence Surveillance and Reconnaissance\\
JIE & Joint Information Environment\\
LPD & Low Probability of Detection\\
LPI & Low Probability of Interception\\
M2M & Machine-to-Machine\\
\end{tabular}

 \noindent\hspace{-0.65em}\begin{tabular}{ll}
MANET & Mobile ad-hoc networks\\
MPE & Mission Partner Environment\\
NCW & Network Centric Warfare\\
NFC & Near Field Communication\\
NFV & Network Function Virtualization\\
PaaS & Platform as a Service\\
PPDR & Public Protection Disaster Relief\\
QoI & Quality of Information\\
RFID & Radio-frequency identification\\
SaaS & Software as a Service\\
SDR & Software Defined Radio\\
SOA & Service-Oriented Architecture\\
TLS & Transport Layer Security\\
UAV & Unmanned Aerial Vehicle\\
VoI & Value of Information\\
Wi-Fi & Wireless Fidelity\\
WSN & Wireless Sensor Networks\\

 \end{tabular}}

%=================================================================
% References: Variant A
%=================================================================
% Back Matter (References and Notes)
%----------------------------------------------------------
% Style and layout of the references
\bibliographystyle{mdpi}

%=====================================
% References, variant A: internal bibliography
%=====================================
\renewcommand\bibname{References}

\end{document}